\documentclass[11pt]{article}
\usepackage{graphicx,comment}
\usepackage{amsmath, amssymb, braket, dsfont,authblk,mathrsfs,etoolbox}
\usepackage{graphicx}
\usepackage{rotating, mathtools}
\usepackage[matrix,arrow]{xy}
\usepackage[numbers]{natbib}
\usepackage[colorlinks]{hyperref}

\thicklines

\usepackage{rotating, mathtools}
\usepackage[matrix,arrow]{xy}
\usepackage[numbers]{natbib}
\usepackage[colorlinks]{hyperref}
\hypersetup{linkcolor={blue}}

\newcommand{\be}{\begin{align}}
\newcommand{\ee}{\end{align}}

\newcommand{\mcf}{\mathcal{F}}

\newcommand{\mcc}{\mathcal{C}}

\newcommand{\mcv}{\mathcal{V}}

\begin{document}

\title{\vspace{-1cm} Integrability and braided tensor categories}
\author{Paul Fendley}
\affil{
{\small All Souls College and Rudolf Peierls Centre for Theoretical Physics,\\ University of Oxford,
Clarendon Laboratory, Oxford OX1 3PU, United Kingdom\vspace{-.3cm}}\\
}

\smallskip 


\maketitle

\begin{abstract} 


Many integrable statistical mechanical models possess a fractional-spin conserved current. Such currents have been constructed by utilising quantum-group algebras and ideas from ``discrete holomorphicity''. I find them naturally and much more generally using a braided tensor category, a topological structure arising in knot invariants, anyons and conformal field theory.  I derive a simple constraint on the Boltzmann weights admitting a conserved current, generalising one found using quantum-group algebras. The resulting trigonometric weights are typically those of a critical integrable lattice model, so the method here gives a linear way of ``Baxterising'', i.e.\ building a solution of the Yang-Baxter equation out of topological data. It also illuminates why many models do not admit a solution. I discuss many examples in geometric and local models, including (perhaps) a new solution.

\end{abstract} 
 
\bigskip

\section{Introduction}

The connection of lattice statistical mechanics to topological invariants has long been known, but perhaps not as well appreciated as it ought to be.  Classic work of the '70s showed how the generators of the transfer matrix satisfied an algebra also having a graphical presentation. 
The canonical example is that of Temperley and Lieb, who related the Potts and six-vertex models by utilising an algebra bearing their name \cite{Temperley1971}. Simultaneously, a geometric expansion for the Potts model was developed where the Boltzmann weight of each term depends on the number of loops in it \cite{Fortuin1971}. The two approaches were unified by showing that generators of the Temperley-Lieb algebra also can be written in terms of loops, where the weight per loop is built into the algebra \cite{Baxter1976}.  A host of other lattice models subsequently were written in the same fashion \cite{Andrews1984,Pasquier1986}. The algebras themselves were also generalised, most notably to the Birman-Murkami-Wenzl algebra \cite{Birman1989,Murakami1990}, providing the analogous setting for more complicated lattice height models already known \cite{Jimbo1988}.

A beautiful manifestation of this connection is in knot and link invariants \cite{Jones1987}.  The Temperley-Lieb algebra underlies the construction of the Jones polynomial, with the loops resulting from resolving the crossings resulting from projecting the loops into two dimensions \cite{Kauffman1991}.  The mathematical structure needed in general is a {\em braided tensor category}.
These categories arose when studying the braiding and fusing of operators in rational conformal field theory \cite{Moore1989}, and now are widely used in physics in the study of anyons \cite{Kitaev2006,Bondersonthesis}. Such categories give not only braid-group representations, but generalise algebras such as Temperley-Lieb to a larger set of rules that give linear relations between the isotopy invariants of labelled graphs. Defining lattice models in terms of a category makes possible finding large classes of topological defects \cite{Aasen2020}.

The connection between statistical mechanics and topology deepens when considering integrable lattice models. Boltzmann weights in an integrable model in two dimensions satisfy a trilinear relation, the {\em Yang-Baxter equation}.  An essential ingredient is expressing its solutions in terms  of  a ``spectral parameter'', enabling the construction of local conserved quantities \cite{Baxter1982}. A key constraint on braiding (the third Reidemeister move) is also trilinear, and indeed, a representation of the braid group and hence a knot/link invariant often can be found by taking an extreme limit of the spectral parameter in a critical integrable lattice model \cite{Jones1989,Wadati1989,Wu1992}. 

Jones coined a term for the converse: to ``Baxterise'' is to construct a $u$-dependent solution of the Yang-Baxter equation starting with the braid-group representation arising in a knot polynomial \cite{Jones1990,Jones2003}. Such an approach generalises the successes starting with the ``universal $R$-matrix'' giving representations of the braid group \cite{Drinfeld1985,Kirillov1990,Khoroshkin1991} found using quantum-group algebras \cite{Gomez1996}. An approach for Baxterising them was initiated by Jimbo \cite{Jimbo1986}, culminating in a formula for the key coefficients giving Boltzmann weights of trigonometric type in many examples \cite{Zhang1990,Delius1994,Delius1995}. Intriguingly, this formula ends up using very little of quantum-group representation theory, despite its rather complicated origin. The only data needed in the end are some tensor-product decompositions and quadratic Casimirs of ordinary simple Lie algebra representations. 

This observation strongly suggests the existence of a more general and more direct way of deriving the trigonometric Boltzmann weights of critical integrable models. The central result of this paper is to provide such a method, and to describe a much simpler approach to Baxterisation. I both simplify and extend the quantum-group results by exploiting the fact that the existence of a {\em conserved current} requires a linear constraint on the Boltzmann weights \cite{Bernard1991}.  These currents have properties suggesting they are lattice analogs of operators in conformal field theory, in particular their behaviour under twists. Indeed, the conserved-current relation gives a set of equations also studied under the guise of  ``discrete holomorphicity'' \cite{Smirnov2006}.  Remarkably, all Boltzmann weights admitting a conserved current that were found in this context are also trigonometric solutions of the Yang-Baxter equation \cite{Cardy2009}. Current conservation thus seems to give a {\em linear} method of Baxterisation. 

I  reformulate and generalise this linear constraint using a braided tensor category, and derive a simple set of conditions guaranteeing a solution. This result extends the construction of conserved currents to a much larger class of models. Even more strikingly, it points to a method to classify which models admit a solution of the Yang-Baxter equation and which do not. Although I have no proof that a fractional-spin conserved current 
 always implies the even stronger constraints coming from the local conserved charges of integrability, it seems a very understandable consequence. Indeed, a standard argument for integrability using the scattering-matrix approach to field theory goes precisely along these lines, albeit with integer-spin charges; see e.g.\ \cite{Shankar77,Dorey97}.



Two classes of lattice models are discussed in this paper: geometric models and local height models. Both types are built from fusion categories, as reviewed in detail in \cite{Aasen2020}. Braiding requires additional data, and is needed here to define and analyse the currents. In geometric models, the Boltzmann weights are expressed directly in terms of the evaluation of fusion graphs, and so are non-local. The most famous example of such are models of self-avoiding loops, where the weight depends on the number of loops present. In height models, the weights are local but the ensuing partition functions are related to those of geometric models via the  ``shadow world'' construction.  Examples include the Ising, Potts, hard-hexagon and self-dual eight-vertex models. Finding the conserved currents in local models by brute force is typically impossible, but the shadow-world construction relates them to the geometric models. The conserved current then in height models follows instantly.

In section \ref{sec:lattice} I give a lightning review of how to build lattice models using fusion categories, compressing the review sections of \cite{Aasen2020} into a few pages. I also review the Yang-Baxter equation in its full spectral-parameter-dependent glory.  Braiding, the currents, and their conservation law are described in section \ref{sec:currents}. Section \ref{sec:central} contains the central result, the formula for the trigonometric Boltzmann weights resulting from requiring they admit a conserved current. Many examples are given in section \ref{sec:examples}, including one not involving a quantum group and one which may be new. The final section \ref{sec:conclusion} contains conclusions and pontifications.


\section{Lattice models from categories}
\label{sec:lattice}

The lattice models at the heart of integrability can be defined in terms of a {\em fusion category}. Fusion categories provide a method of defining and computing topological invariants of labelled trivalent graphs on the plane.
In this section I describe key facts about fusion categories, and how to define integrable geometric lattice models in terms of them. Geometric models have non-local Boltzmann weights, but provide the most transparent way of starting the analysis. Moreover, the category structure makes the translation of these results to locally defined height models straightforward.

\subsection{Evaluating graphs using fusion categories}
\label{sec:fusion}

Here I summarize the key background needed to build the currents.  Much more detailed reviews of tensor categories can be found in \cite{Moore1989,Bakalov2001,Kitaev2006,Walker2006}.

The core of a fusion category is a finite set of {\em simple objects}. Objects form a vector space, with simple objects a basis.  Examples of simple objects include the primary fields in a rational conformal field theory and anyon types in a 2+1d system with topological order.  The {\em fusion algebra} governs the tensor product of simple objects $a$ and $b$:
\begin{align}
a\otimes b = \bigoplus_c N^c_{ab}\ c\ ,
\label{Ndef}
\end{align}
with the non-negative integers $N_{ab}^c$ describing the sum over simple objects $c$. The irreducible representations of a semi-simple Lie algebra obey a fusion algebra, but since there are an infinite number of them, do not make up a fusion category. However, deforming the Lie algebra to a quantum-group algebra can truncate the allowed representations to yield simple objects obeying \eqref{Ndef}.

A fusion category $\mcc$ gives a method for associating an isotopy invariant to labelled planar trivalent graph called a {\em fusion diagram} $\mathcal{F}$.  Each edge is labelled by a simple object in the category, and the label on each edge touching a trivalent vertex must have $N_{ab}^c\ne 0$.  The invariant associated to each $\mathcal{F}$ is called the {\em evaluation}, and denoted eval${}_\mcc[\mcf]$. Invariance under isotopy means that the evaluation remains the same under any continuous deformation of the fusion diagram preserving the labels. 

All fusion categories have an identity object $0$, obeying $a\otimes 0=0\otimes a = a$ for all $a$.  Labelling an edge by $0$ is equivalent to omitting that edge. For simplicity I mainly discuss only self-dual objects, which have the identity $0\in a\otimes a$. Self-duality means that one does not need to include arrows on the edges of the diagram. In addition, I also make the simplifying assumptions described in sec.\ 2 of \cite{Aasen2020}, for example considering only $N_{ab}^c=N_{ba}^c=0,1$. 

Two sets of linear identities allow fusion diagrams to be evaluated. $F$ {\em moves} relate graphs as
\begin{align}
\mathord{\vcenter{\hbox{\includegraphics[scale=0.45]{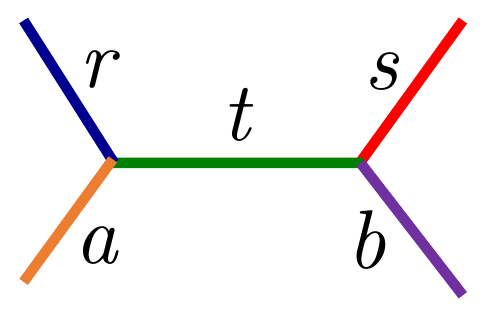}}}}\quad =\ \sum_{t'} F_{tt'}\begin{bmatrix} r&s\\a&b\end{bmatrix}\ 
\mathord{\vcenter{\hbox{\includegraphics[scale=0.45]{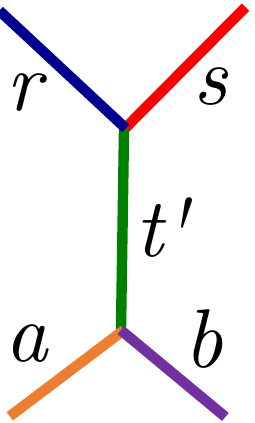}}}}
\label{Fmove}
\end{align}
where the coefficients are called $F$ {\em symbols}. This identity relates two graphs by replacing the subgraph on the left with that on the right, leaving the rest of the diagram unchanged. 
Since the labels of the external lines $a,b,r,s$ in \eqref{Fmove} do not change in the $F$ move, one can think of the symbols as matrices acting on the internal labels $t$ and $t'$. 
The one additional ingredient needed to evaluate a planar fusion diagram is ``bubble removal'', which relates diagrams as
\begin{align}
\mathord{\vcenter{\hbox{\includegraphics[scale=0.4]{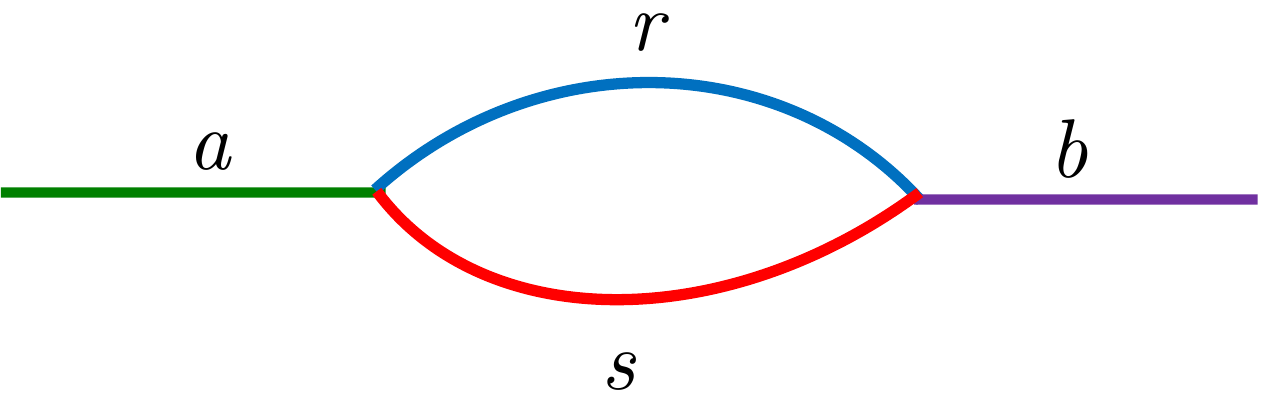}}}}\quad =\ \delta_{ab}\,  \ \sqrt\frac{d_r d_s}{d_a}\;
\mathord{\vcenter{\hbox{\includegraphics[scale=0.4]{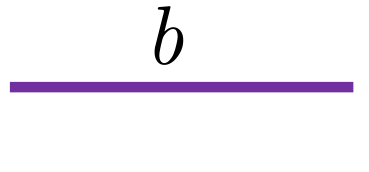}}}}\ .
\label{bubbleremoval}
\end{align}
The number $d_a$ associated with each simple object is called the {\em quantum dimension}, and like the $F$ symbols is specified by the category. Namely, each $d_a$ is the largest eigenvalue of the matrix $\hat{N}_a$ that has entries $(\hat{N}_a)_b^c= N_{ab}^c$. Thus $d_0=1$, and more generally
\begin{align}
d_a d_b = \sum_c N_{ab}^c d_c\ .
\label{dabc}
\end{align} 
The nicest way of understanding the meaning of the quantum dimension comes from setting $a=b=0$ in \eqref{bubbleremoval}, showing that evaluating a single closed loop labeled by $a$ gives $d_a$. Setting $a=0$ but having $b\ne 0$ means that ``tadpoles'' have vanishing evaluation.

Given  \eqref{Fmove} and \eqref{bubbleremoval}, one can evaluate any fusion graph simply by using $F$ moves to make a bubble, removing it, and then repeating. The beauty of the category is that the evaluation is independent of which order the $F$ moves are done. Eventually, this process yields a sum over loop configurations, evaluated simply with a weight $d_a$ per loop of type $a$. For example, a triangle can be removed by doing an $F$ move and then bubble removal:
\begin{align}
\mathord{\vcenter{\hbox{\includegraphics[scale=0.5]{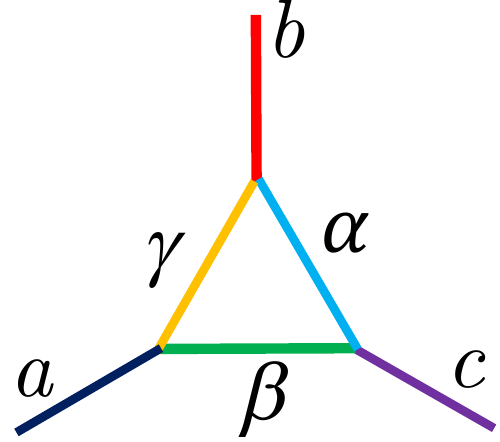}}}}\quad=\  \sum_r
F_{\beta r} \begin{bmatrix} \gamma&\alpha\\ a&c\end{bmatrix}\ 
\mathord{\vcenter{\hbox{\includegraphics[scale=0.45]{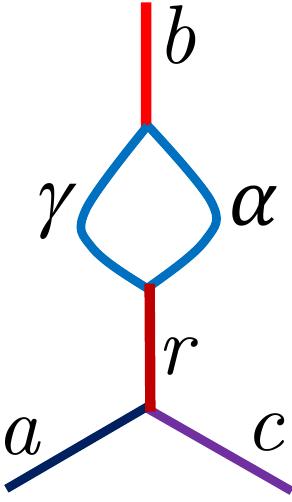}}}}\quad =\ 
\sqrt{\frac{d_\alpha d_\gamma}{d_b}}F_{\beta b} \begin{bmatrix} \gamma&\alpha\\ a&c\end{bmatrix} \ 
\mathord{\vcenter{\hbox{\includegraphics[scale=0.5]{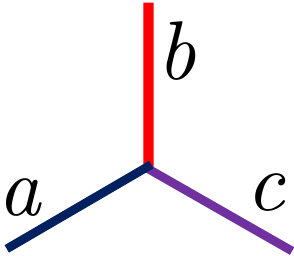}}}}
\label{triangleremove}
\end{align}
For the linear relations \eqref{Fmove} to be consistent, the $F$ symbols must satisfy a variety of constraints, most famously the pentagon identity. Simpler ones follow by using different $F$ moves to simplify \eqref{triangleremove}:
\begin{align}
\sqrt{\frac{d_\alpha d_\gamma}{d_b}}\ F_{\beta b} \begin{bmatrix} \gamma&\alpha\\ a&c\end{bmatrix} \ =
\sqrt{\frac{d_\alpha d_\beta}{d_c}}\ F_{\gamma c} \begin{bmatrix} a&b\\ \beta&\alpha\end{bmatrix} \ =
\sqrt{\frac{d_\gamma d_\beta}{d_a}}\ F_{\alpha a} \begin{bmatrix} b&c\\ \gamma&\beta\end{bmatrix} \ .
\label{usefulid}
\end{align}
Other identities arise when when one of the labels is $0$:
\begin{align}
F^{}_{tt'}\begin{bmatrix} r&0\\a&b\end{bmatrix} = \delta_{tb} \delta_{t'r} N_{ab}^r\ ,\qquad\quad 
F^{}_{s0}\begin{bmatrix} r&r\\a&a\end{bmatrix}=F^{}_{0s}\begin{bmatrix} r&r\\a&a\end{bmatrix} =  \sqrt{\frac{d_s}{d_ad_r}}N_{as}^r\ .
\label{F0}
\end{align}
The former follows simply by omitting the external line labeled $0$, while the latter arises by doing an $F$ move on \eqref{bubbleremoval} and noting that tadpoles vanish.  A useful consequence of the latter identify in \eqref{F0} is that two adjacent strands can be joined via 
\begin{align}
\mathord{\vcenter{\hbox{\includegraphics[scale=0.45]{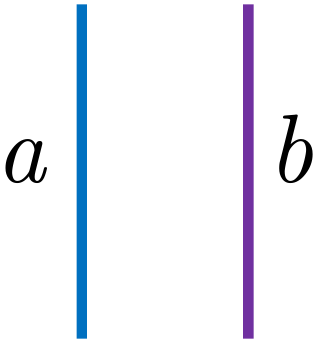}}}}\quad =\ \sum_{\chi} \sqrt{\frac{d_\chi}{d_ad_b}}\ \mathord{\vcenter{\hbox{\includegraphics[scale=0.45]{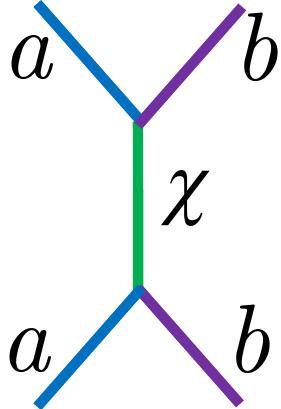}}}}\ .
\label{joinstrands}
\end{align}

 
\subsection{Geometric models}
\label{sec:geometric}

The degrees of freedom in geometric models are expressed in terms of fusion diagrams. It thus seems most natural to define the models on a honeycomb lattice, with each edge labelled by some object in the category so that each trivalent vertex can be related to the fusion of these objects. However, integrable models are best dealt with on the square lattice. Thus I start out with the latter on the plane, with periodic boundary conditions in the horizontal direction and open in the vertical. 

The first step in defining a geometric model is labelling by some object $\rho \in \mathcal{C}$ each edge of the square lattice. 
The next step is to crack open each vertex into two trivalent vertices connected by a horizontal line segment as
\begin{align}
 \mathord{\vcenter{\hbox{\includegraphics[scale=0.45]{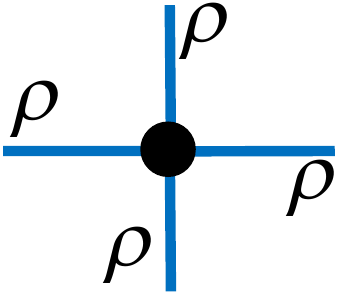}}}}\ = \sum_{\chi} \frac{\sqrt{d_\chi}}{d_\rho}\,A_\chi \ \mathord{\vcenter{\hbox{\includegraphics[scale=0.45]{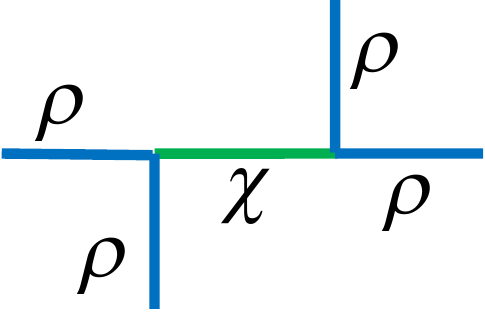}}}}\ .
 \label{crackopen}
\end{align} 
The additional horizontal line is labelled by an object $\chi\in \rho\otimes\rho$. The amplitudes $A_\chi$ are complex numbers not determined by the category data. The explicit quantum dimensions included look unwieldy here, but turn out to be a useful convention.  The cracking open turns the square lattice into a brick lattice, topologically equivalent to the honeycomb lattice. 

A {\em completely packed} geometric model corresponds to taking  $\rho$ to be a simple object. The degrees of freedom $\chi_{\rm v}$ then live only on the vertices ${\rm v}$ of the square lattice (the edges of the brick lattice that do not belong to the square lattice).
 Each labelling $\{\chi_{\rm v}\}$ defines a planar fusion diagram $\mcf$ on the brick lattice, and the partition function is equivalently the sum over all allowed $\mcf$. 
The Boltzmann weight for each $\mcf$ has both local and non-local parts. The non-local part is simply the evaluation of corresponding fusion diagram, and so depends only on its topology.  The local part is expressed in terms of amplitudes $A_{\chi_{\rm v}}$, each of which depends only on the label $\chi_{\rm v}$ on the corresponding edge in $\mcf$.  With these definitions the partition function of a completely packed geometric model is
\begin{align}
Z_\rho=\sum_{\mcf} {\rm eval}_{\mathcal{C}}[{\mcf}]\, \prod_{\rm v}   \frac{\sqrt{d_{\chi^{}_{\rm v}}}}{d_\rho}A_{\chi_{\rm v}^{}}\ .
\label{Zgeom}
\end{align}
A picture for the partition function can be written using \eqref{crackopen} as
\begin{align}
Z_\rho =\hbox{ eval}\  \mathord{\vcenter{\hbox{\includegraphics[scale=0.45]{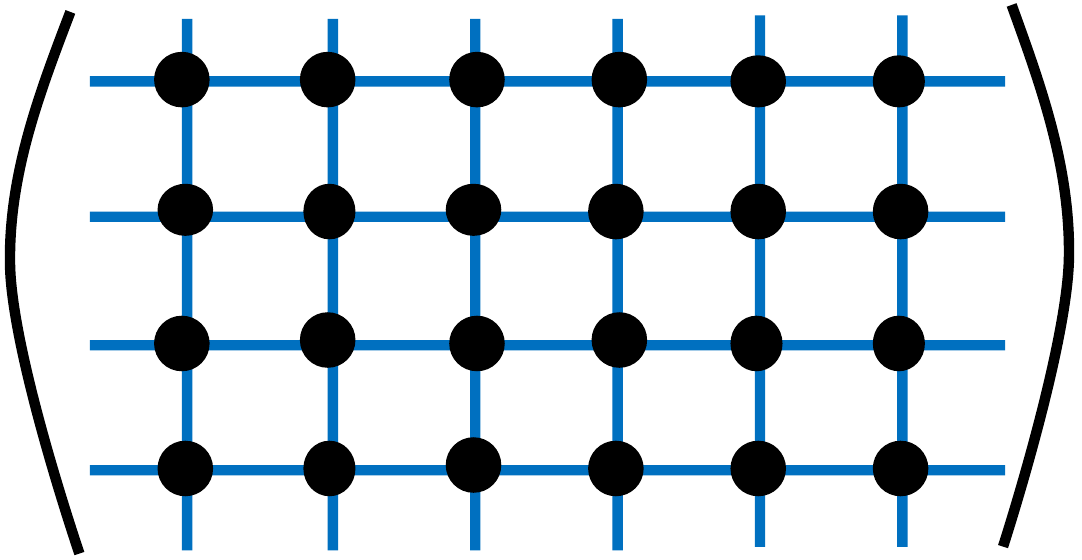}}}}\ .
\label{Zpicture}
\end{align}
Unlabelled solid lines in this paper are always $\rho$ lines.  More general geometric models come by relaxing the requirement that $\rho$ be simple, so that $\rho=\oplus_j \rho_j$ for simple $\rho_j$. In such models the degrees of freedom live on the edges as well.


The simplest and most famous examples of geometric models are {\em loop models}, where the fusion diagrams can be rewritten in terms of self-avoiding loops. Loop models arise from categories when there are only two channels $\chi=0,1$ allowed on each face, i.e.\  $\rho\otimes\rho = 0 + 1$. Using \eqref{crackopen} gives configurations including trivalent vertices with labels $\rho,\rho,1$. However, each such fusion diagram can be turned into a sum over loops by exploiting the $F$ moves. The needed relation comes from \eqref{F0}:
\begin{align}
\mathord{\vcenter{\hbox{\includegraphics[scale=0.45]{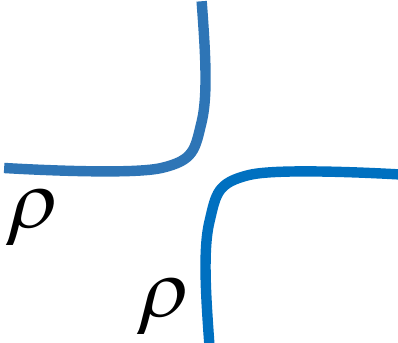}}}}\quad=\ \frac{\sqrt{d_1}}{d_\rho}\  \mathord{\vcenter{\hbox{\includegraphics[scale=0.45]{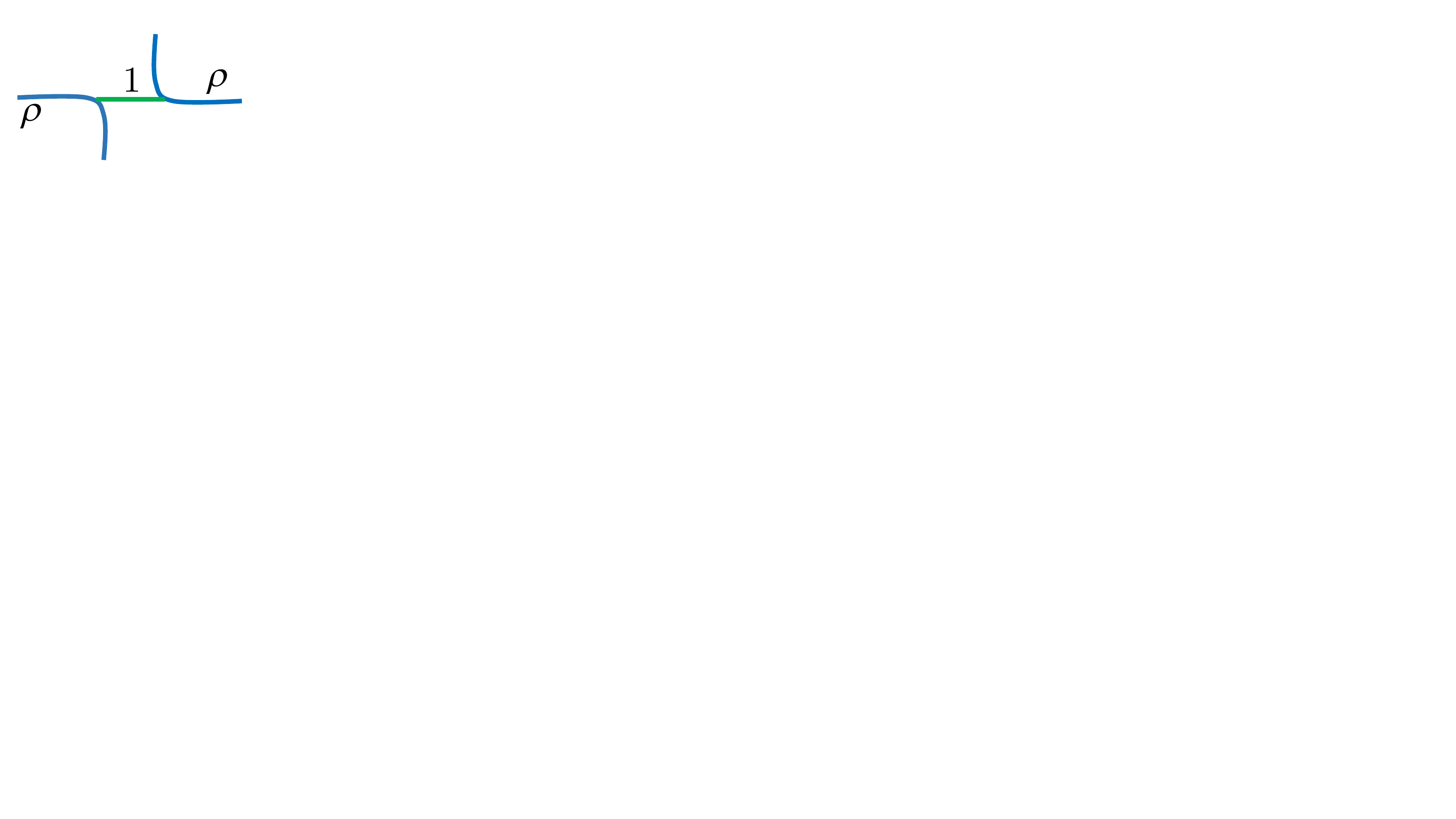}}}}\quad +\ \frac{1}{d_\rho}\ \mathord{\vcenter{\hbox{\includegraphics[scale=0.45]{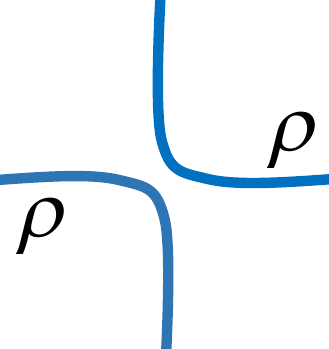}}}}
\label{squigglyremove}
\end{align}
where \eqref{dabc} requires $d_1 = d_\rho^2-1$. Any lines labeled by $1$ can be replaced by are a linear combination of the two
self-avoidances, and so \eqref{crackopen} can be recast as
\begin{align}
 \mathord{\vcenter{\hbox{\includegraphics[scale=0.45]{dualvertex.pdf}}}} =  A_1 \ \mathord{\vcenter{\hbox{\includegraphics[scale=0.45]{loop1.pdf}}}}\quad+\ \frac{A_0  - A_1}{d_\rho}\ \mathord{\vcenter{\hbox{\includegraphics[scale=0.45]{loop2.pdf}}}}\ .
\label{recast}
\end{align}

Using \eqref{recast} at each vertex turns a sum over fusion diagrams into a sum over completely packed self-avoiding loops $\mathcal{L}$. The local weights for each loop configuration are determined by the coefficients in \eqref{recast}. Moreover, since this rewriting uses an $F$ move, it does not change the evaluation. Evaluation is now easy to do, as each closed loop yields a weight $d_\rho$ per loop, giving the partition function
\begin{align}
Z_{\rm CPL} = \sum_{\mathcal{L}} 
\big(d_\rho\big)^{n_{\rm L}-n_0} (A_0-A_1)^{n_0} (A_1)^{n_{\widehat 0}}\  ,
\label{ZCPL}
\end{align}
where $n_{\rm L}$ is the number of loops and $n_0,\, n_{\hat 0}$ count each type of avoidance in $\mathcal{L}$.
The model is often called an $O(N)$ loop model for historic reasons, with $N$ the weight per loop.  In the corresponding random-cluster model, the loops surround clusters of weight $Q=d_\rho^2$, which for integer $Q$, it can be mapped onto the (local) $Q$-state Potts model \cite{Fortuin1971,Baxter1982}. Although loops and \eqref{ZCPL} look much simpler than fusion diagrams and \eqref{Zgeom}, the identities stemming from the category make the latter formulation a much better setting for introducing and analysing the currents.



\subsection{Height models}
\label{sec:heights}

By construction, the Boltzmann weights of the geometric model are non-local. A remarkable fact is that for {\em any} geometric model built on a fusion category, there exists a {\em local} model with a related partition function. Simple examples came from \cite{Temperley1971}, where the generators of the transfer matrix were shown to satisfy the Temperley-Lieb algebra. Local and non-local models give rise to different representations, but the algebra provides relations between the ensuing partition functions. Such algebraic results were generalised to the full category setting, under the name of {\em shadow world} \cite{Reshetikhin1988,Turaev1992,TuraevViro,Barrett1996}.
The  local models go under a variety of names, with RSOS (restricted-solid-on-solid), IRF (interactions round a face),  and anyon chains among them. I call them {\em height models.}

The heights are objects in the category $\mcc$ living on the dual square lattice, the faces of \eqref{Zpicture}. They satisfy {\em adjacency rules} dictated by fusion rules coming from $\rho$, the object used to define the geometric model.  These rules are conveniently displayed in {\em fusion trees}, fusion diagrams of the form
\begin{align}
 \mathord{\vcenter{\hbox{\includegraphics[scale=0.45]{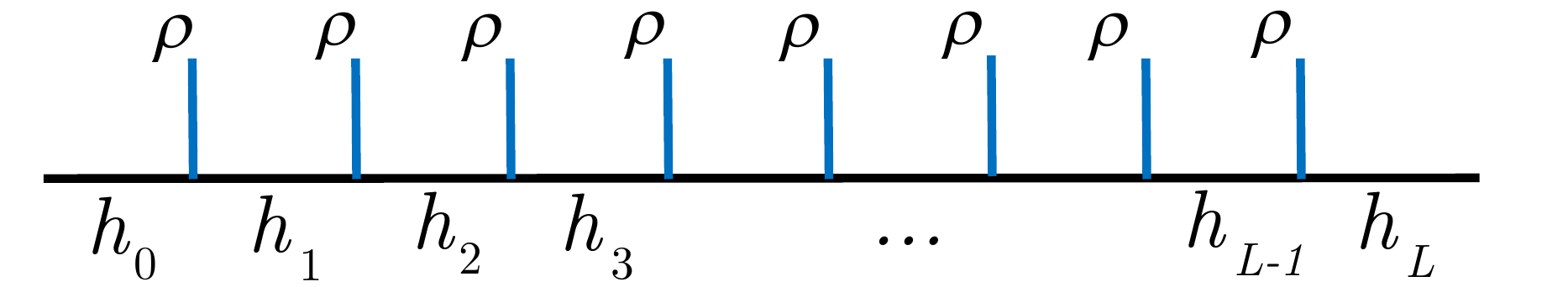}}}}
\label{fusiontree}
\end{align}
The $L$ vertical lines are all labeled with $\rho$, and I call them {\em strands}. The {heights} $h_j$ with $j=0,\dots L$ are the horizontal lines, and by the fusion rules must satisfy $h_{j\pm 1}\in \rho\otimes h_{j}$. For $\rho$ simple, this condition translates to $N_{\rho h_{j}}^{h_{j\pm 1}} >0$.  Each allowed labelling of such a fusion tree corresponds to a height configuration for one row of the lattice. For open boundary conditions, there are $L+1$ heights, while with periodic boundary conditions $h_{0}=h_L$,  giving $L$ of them.

Operators in height models act on a vector space $\mathcal{V}$ whose basis elements are all the allowed fusion trees for a given $\rho$ and $L$. Operators are then defined using a fusion diagram with some number of $\rho$ strands at the bottom and the same number at the top; see e.g.\ \eqref{Pdef} for a two-strand operator.  Such an operator acts on $\mcv$ by gluing it to the tree somewhere. Although perhaps gluing initially yields a more complicated fusion diagram, as long as it does not wrap around a cycle it can always be reduced to a fusion tree \eqref{fusiontree} by doing $F$ moves and bubble removal. The rules of the category guarantee that the results are independent of how this reduction is done. 

The Boltzmann weights are built from two-strand {\em projection operators} 
 \begin{align}
 P^{(\chi)} \equiv \frac{\sqrt{d_\chi}}{d_\rho}\ \mathord{\vcenter{\hbox{\includegraphics[scale=0.4]{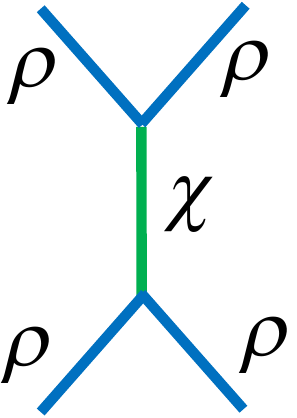}}}}\ .
 \label{Pdef}
 \end{align}
If $\rho$ is not simple, one defines a set of projection operators labeled by the simple objects on the strands. A set of operators $P^{(\chi)}_j$ acting on $\mcv$ is then defined by gluing the two bottom strands in \eqref{Pdef} to the two strands surrounding $h_j$. Using $F$ moves to simplify the resulting picture into a sum over trees of the form \eqref{fusiontree} gives then $P^{(\chi)}_j:\ \mcv\to\mcv$. Explicitly,
\begin{align}
\begin{split}
\frac{\sqrt{d_\chi}} {d_\rho}\mathord{\vcenter{\hbox{\includegraphics[scale=0.45]{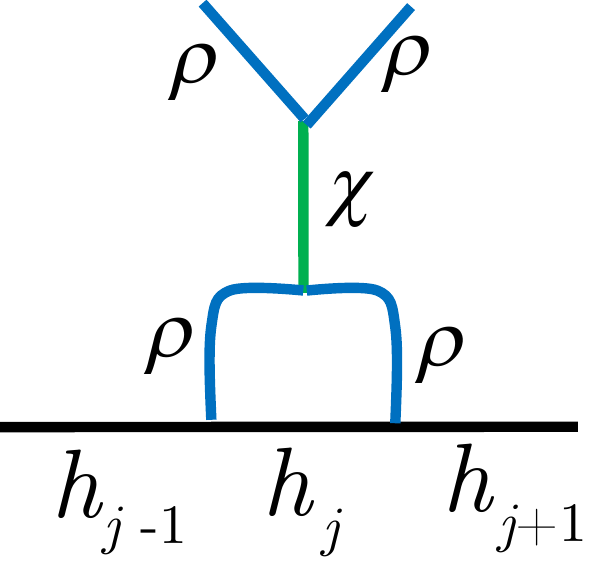}}}}\quad
 &=\ F_{h_j \chi} \begin{bmatrix} \rho&\rho\\ h_{j-1}&h_{j+1}\end{bmatrix}\,\mathord{\vcenter{\hbox{\includegraphics[scale=0.45]{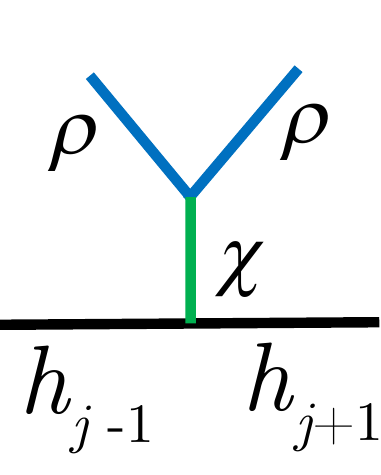}}}} \\[-.5cm]
 &=\ \sum_{h_{j}'}F_{h_j\chi} \begin{bmatrix} \rho&\rho\\ h_{j-1}&h_{j+1}\end{bmatrix}
 F_{\chi h_j'} \begin{bmatrix} h_{j-1}&\rho\\ h_{j+1}&\rho\end{bmatrix}\ \mathord{\vcenter{\hbox{\includegraphics[scale=0.45]{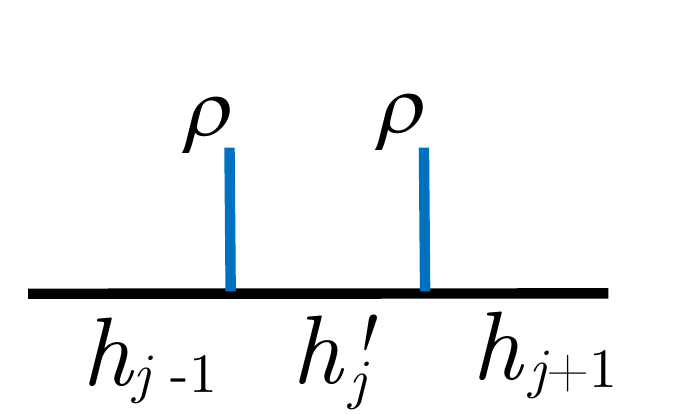}}}}
\end{split}
\label{Pmatrix}
\end{align}
using \eqref{triangleremove} and \eqref{Fmove}.
The matrix elements of each $P_j^{(\chi)}$ in the basis \eqref{fusiontree} are thus
\begin{align}
\Big(P_j^{(\chi)}\Big)_{\{h\},\{h'\}} = F_{h_j\chi} \begin{bmatrix} \rho&\rho\\ h_{j-1}&h_{j+1}\end{bmatrix}
 F_{\chi h_j'} \begin{bmatrix} h_{j-1}&\rho\\ h_{j+1}&\rho\end{bmatrix}\prod_{n\ne j} \delta_{h_{n} h_n'}\ .
\label{Pmatrixelements}
\end{align}

One remarkable feature of the category setup is that the conserved currents in the height models can be defined and analysed without ever needing the explicit expressions \eqref{Pmatrixelements}. Relations among operators are derived by gluing them together and manipulating using the relations between fusion diagrams. For example, proving that the $P^{(\chi)}_j$ for each $j$ form a complete set of orthogonal projectors is easy. Gluing the top strands of one to the bottom strands of the other and using  \eqref{bubbleremoval} gives
\begin{align}
\mathord{\vcenter{\hbox{\includegraphics[scale=0.4]{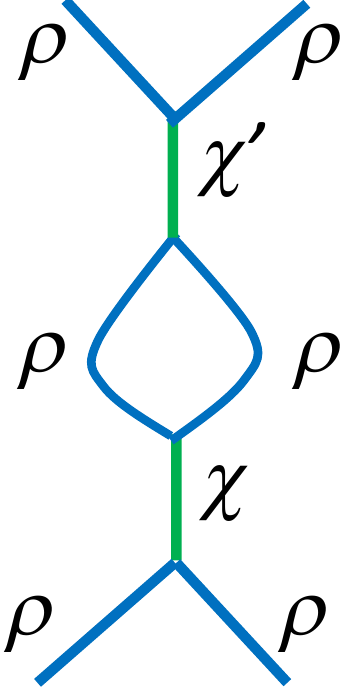}}}}\quad =\ \delta_{\chi\chi'} \frac{d_\rho}{\sqrt{d_\chi}}\ \mathord{\vcenter{\hbox{\includegraphics[scale=0.45]{projector.pdf}}}}\qquad\quad\implies\quad P_j^{(\chi)}P_j^{(\chi')}=\delta_{\chi\chi'} P_j^{(\chi)}\ .
\label{projalg}
\end{align}
Setting $a=b=\rho$ in \eqref{joinstrands} shows they sum to the identity operator: $\sum_\chi P_j^{(\chi)}=\mathds{1}$. In a particular category for particular choices of $\rho$,  relations between projectors at different $j$, can then be derived, obtaining algebras such those of Temperley-Lieb \cite{Temperley1971} or the Birman-Wenzl-Murakami \cite{Birman1989,Murakami1990}.

Comparing \eqref{Pdef} to \eqref{crackopen} shows that the cracking-open process used to define the geometric models amounts to a sum over projection operators. Related height models are then defined via
\begin{align}
R_j \equiv \sum_{\chi\in\rho\otimes\rho} A_\chi P_j^{(\chi)}
\label{RP}
\end{align}
for the same amplitudes $A_\chi$ (note the explicit quantum dimensions in \eqref{crackopen} normalize the amplitudes to multiply a projector).  If $\rho$ is not simple, then the appropriate edge labels $\nu_{j}$ must be included in the amplitudes. Acting with this $R$ operator can be thought of as adding a vertex to the square lattice. For periodic boundary conditions, the basis elements of $\mcv$ are labeled by all the allowed height configurations with $h_0=h_L$. The transfer matrix acting on this $\mathcal{V}$ is then
\begin{align}
T= R_L R_{L-1} \dots R_2 R_1\ .
\end{align}
This $T$ acts at a 45-degree angle to the square lattice, i.e.\ across the diagonals of the squares.

With appropriate choices of boundary conditions, the partition functions of the geometric models and the height models can then be related, as follows from the shadow-world construction \cite{Reshetikhin1988,Turaev1992,TuraevViro,Barrett1996}. For example, the ``restricted solid-on-solid'' models of Andrews, Baxter and Forrester  \cite{Andrews1984} are related to the completely packed loop models in this fashion. The quantum spin chains found by taking the Hamiltonian limit of the transfer matrix have been studied in the guise of ``anyon chains'' \cite{Feiguin2007}. Other connections of local models to associated categories are described in detail in \cite{Aasen2020}.

\subsection{The Yang-Baxter equation}
\label{sec:YBE}

The purpose of this paper is to find linear equations for Boltzmann weights that ``Baxterise'' \cite{Jones1990} the braided tensor category, i.e.\ give solutions of the much-more complicated Yang-Baxter equation.  The latter is never needed for the analysis, but I give it here both for the sake of completeness and to provide some intuition into how to parametrise its solutions. 

The Yang-Baxter equation (YBE) is trilinear in the Boltzmann weights. A one-parameter family of commuting transfer matrices can be constructed using its solutions. This parameter is typically called the {\em spectral parameter}, as the eigenvalues of the transfer matrix depend on it even though the eigenvectors do not. Many of the most profound results of integrability come from analysing how physical quantities depend on the spectral parameter \cite{Baxter1982}. Mathematical ones do too: as reviewed in section \ref{sec:braids}, braid-group generators often can be found by taking an extreme limit. 
In a geometric model, only the {\em local} part of the Boltzmann weights depend on the spectral parameter $u$, and the evaluation of any individual fusion diagram is not affected by its presence. I thus sometimes write the amplitudes as $A_\chi(u)$, but these amplitudes may very well depend on other parameters. 

The three Boltzmann weights in each term of the YBE have distinct spectral parameter, but the YBE requires a relation between them, and so is a two-parameter equation.  Labelling each Boltzmann weight in  \eqref{crackopen} by the corresponding spectral parameter, the pictorial version of the YBE is
\begin{align}
 \mathord{\vcenter{\hbox{\includegraphics[scale=0.45]{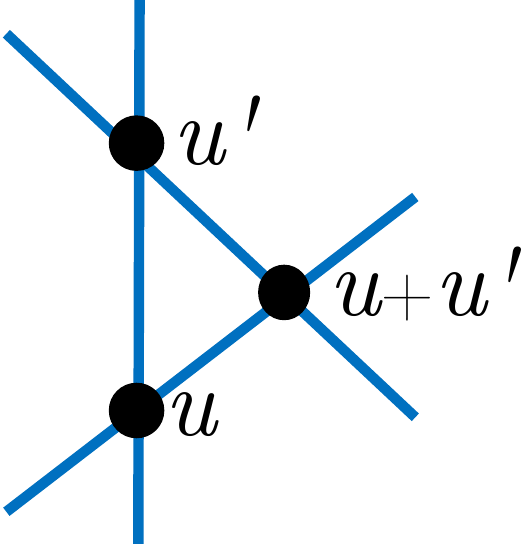}}}}\quad   =  \quad\mathord{\vcenter{\hbox{\includegraphics[scale=0.45]{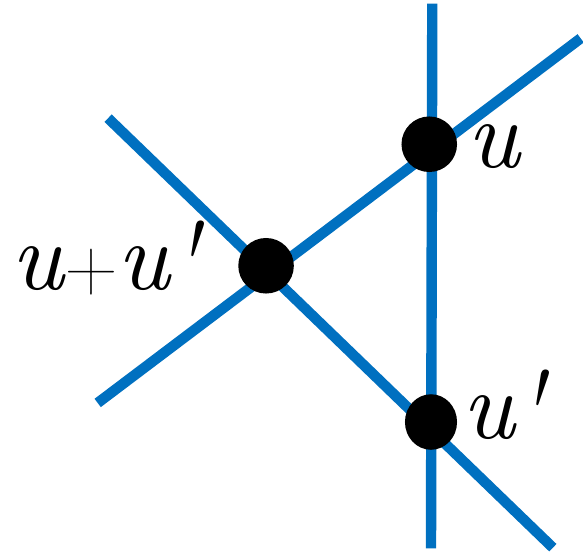}}}} 
\label{YBE}
\end{align}
The YBE applies to the height models when the Boltzmann weights are treated as a two-strand operator defined in \eqref{RP}, giving a three-strand relation
\begin{align}
R_{j}(u)\, R_{j+1}(u+u')\, R_j(u') = 
R_{j+1}(u')\, R_{j}(u+u')\, R_{j+1}(u)\ .
\label{YBE2}
\end{align}
The relation between the parameters given in  (\ref{YBE}\,\ref{YBE2})  is of ``difference form'', as all the solutions discussed in this paper have this property. This form can be generalised and solutions found \cite{AuYang1987}, but I defer the discussion of the related conserved currents to the future. One nice feature of the difference form is that each spectral parameter in each Boltzmann weight can be interpreted as the (bottom) {\em angle} between those two lines: The relation between the three spectral parameters in   (\ref{YBE}\,\ref{YBE2}) is required for the picture to lie in the plane.

Solving the YBE even in simple cases requires work. One plugs in the expansion in \eqref{crackopen}, and then can deform, use $F$ moves and bubble removal to relate different fusion diagrams for the three strands. The equality in \eqref{YBE} must then hold for each of a linearly independent set of fusion diagrams, (over) constraining the amplitudes. Since the corresponding operators $R_j$ from \eqref{RP} in the height models are written in terms of the same projectors, the same amplitudes give a solution to \eqref{YBE2} as well.

The completely packed loop model provides a nice illustration of the techniques. It is easiest to use the loop basis for the Boltzmann weights \eqref{recast}, as the set of linearly independent fusion diagrams is more apparent. These diagrams are
\begin{align}
\mathord{\vcenter{\hbox{\includegraphics[scale=0.45]{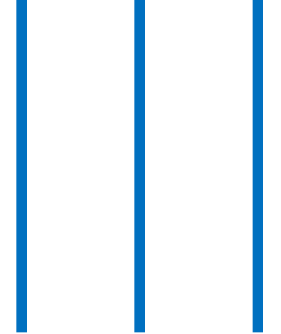}}}}\ ,\qquad\   
\mathord{\vcenter{\hbox{\includegraphics[scale=0.45]{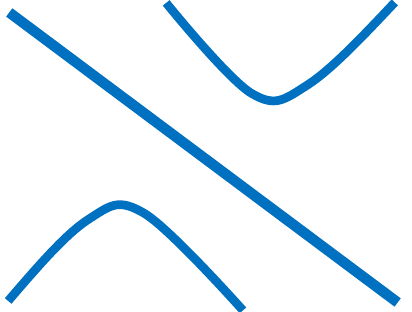}}}}\ ,\qquad\   
\mathord{\vcenter{\hbox{\includegraphics[scale=0.45]{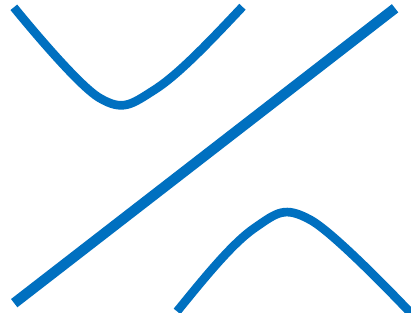}}}}\ ,\qquad\  
\mathord{\vcenter{\hbox{\includegraphics[scale=0.45]{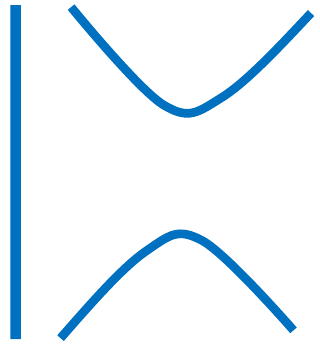}}}}\ ,\qquad\  
\mathord{\vcenter{\hbox{\includegraphics[scale=0.45]{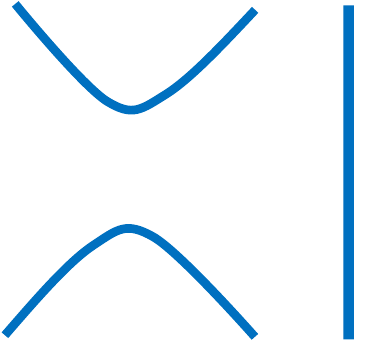}}}}\ 
\label{basisloops}
\end{align}
After plugging in \eqref{recast} and simplifying, the YBE \eqref{YBE} gives a relation for the coefficients of each of the fusion diagrams in \eqref{basisloops}. Namely, letting $C=(A_0 - A_1)/d_\rho$,  the first diagram yields
\[
A_1(u) A_1(u+u') A_1(u') = A_1(u') A_1(u+u') A_1(u)\ 
\]
while the second and third each give
\[
C(u) C(u+u')A_1(u) = A_1(u') C(u+u') C(u)\ .
\]
These three are all automatically satisfied. The relation for the fourth picture is
\begin{align}
A_1(u)C(u+u')A_1(u') =& d_\rho\, C(u') A_1(u+u') C(u)
+ A_1(u') A_1(u+u')C(u)\cr
&\qquad + C(u') A_1(u+u') A_1(u) + C(u)C(u+u')C(u')
\label{YBEloopeqn}
\end{align}
with the same equality for the fifth. The factor $d_\rho$ arises by removing a closed loop. 

Demanding the Boltzmann weights in the completely packed loop model satisfy the YBE thus yields a nasty-looking non-linear functional equation \eqref{YBEloopeqn} for the amplitude ratio. The solution, however, is remarkably simple. Defining the parameter $q$ via $d_\rho = q+q^{-1}$ gives
\begin{align}
\frac{C(u)}{A_1(u)} = \frac{e^{u} -1}{q-q^{-1}e^{u}}\quad \implies \quad
\frac{A_0(u)}{A_1(u)}  = \frac{e^{u}q -q^{-1}}{q-q^{-1}e^{u}}\ .
\label{YBEloop}
\end{align}

\section{Conserved currents}
\label{sec:currents}

Defining the lattice models in terms of category data as in section \ref{sec:lattice} leads to a variety of useful applications. In particular, topological defects can be constructed in any lattice model built from a fusion category \cite{Aasen2020}. Deforming the path of a topological defect leaves the corresponding partition function invariant. Even more remarkably, such defects are defined so that they branch and fuse in a topologically invariant fashion, leaving the partition function invariant under the $F$ moves of the category. Exploiting these properties allows certain universal quantities to be computed directly and exactly on the lattice \cite{Aasen2016,Aasen2020}. 

The current $J(z)$ is defined by {\em terminating} a topological defect in a {\em non-topological} fashion at a location $z$. Correlation functions of such operators then will depend on $z$. The current is non-local because of the topological defect emanating from it, but in a very gentle way, as the path can be deformed without changing the correlator.  A conserved current here satisfies a lattice version of a divergence-free condition,  originally introduced in models with quantum-group symmetries \cite{Bernard1991}. This condition was reintroduced and rebranded as ``discrete holomorphicity'' \cite{Smirnov2006}, but as I will explain, calling it a conserved-current relation is more appropriate.   

In this section I define the conserved currents in terms of a {\em braided tensor category}. This type of category contains more data than the fusion category used to define the lattice models studied here and the topological defects studied in \cite{Aasen2016,Aasen2020}. The construction here is simpler, but less general. The payoff is that not only does a braided tensor category give a natural definition for the currents, but using its rules gives a simple method for finding Boltzmann weights where they are conserved.



\subsection{Braiding and categories}
\label{sec:braids}

While the lattice models themselves can be defined using a fusion category, the simplest way to construct conserved currents is to include some additional structure, {\em braiding}. One nice way to think about braiding is to imagine a knot or link in three dimensions, and then projecting it onto the plane. The projection results in overcrossing and undercrossings, drawn respectively as 
\begin{align}
B^{(bc)}\equiv \mathord{\vcenter{\hbox{\includegraphics[scale=0.45]{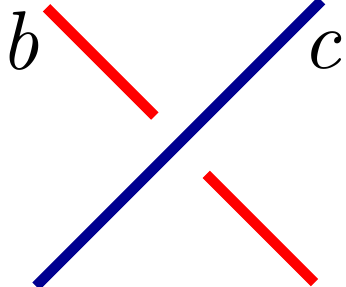}}}} ,\qquad\qquad   
\overline{B}^{(bc)}\equiv {\mathord{\vcenter{\hbox{\includegraphics[scale=0.45]{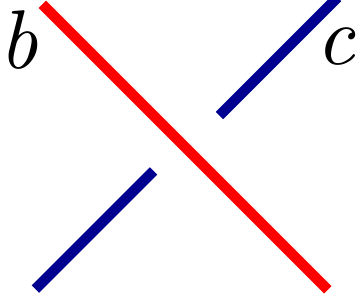}}}}}\ ,\qquad\   
\label{braidpicture}
\end{align}
The two are oriented with respect to the square lattice on which the lattice models are defined.

The braiding must satisfy various consistency relations, known as  {\em Reidemeister moves}, to ensure that the resulting topological invariant is independent of projection. Two of them ensure that the crossings in \eqref{braidpicture} are generators of the {\em braid group} \cite{Birman1975}. The group generators $B^{(bc)}_{j}$ and $\overline{B}^{(bc)}_{j}$ act non-trivially on two strands $j,j+1$, with multiplication gluing one set of ends together. The second Reidemeister move is the fact that the two pictures in \eqref{braidpicture} are inverses:  
$B^{(bc)}_{j} \overline{B}^{(cb)}_{j} = \mathds{1}$. 
The {third Reidemeister move} involves three strands, and is
\begin{align}
B_j B_{j+1} B_j\ =\  \mathord{\vcenter{\hbox{\includegraphics[scale=0.45]{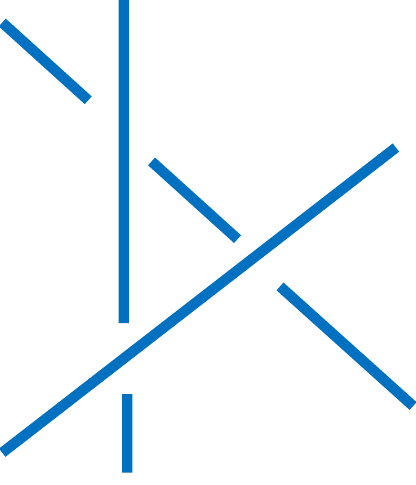}}}}\ =
 \mathord{\vcenter{\hbox{\includegraphics[scale=0.45]{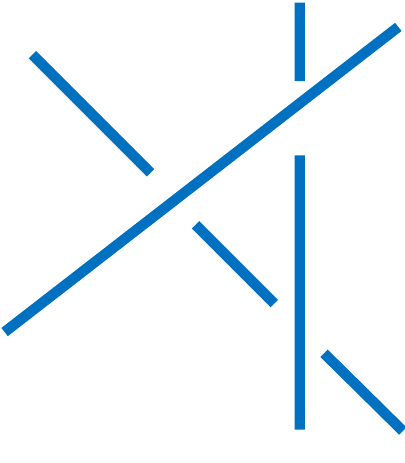}}}}\ =\ B_{j+1}B_jB_{j+1}
\label{Reidemeister3}
\end{align}
where for simplicity all lines are labeled by the same object $\rho$ and the superscripts omitted.

The resemblance of \eqref{Reidemeister3} to \eqref{YBE} is obvious, and often it is referred to as the Yang-Baxter equation in the category literature. However, this name is fairly misleading. Not only does analysis of the braid group long predate both Yang and Baxter\footnote{Artin coined the term in 1925, but there are antecedents \cite{Magnus1974}. Yang was born in 1922, Baxter 1940.}, but it does not include the all-important dependence on the spectral parameters apparent in \eqref{YBE}. 
The resemblance does make it fairly obvious how to obtain representations of the braid group from solutions to the Yang-Baxter equation, as all the arguments in the latter are the same for $u=u'=0$, and $|u|\to \infty$, $|u'|\to\infty$ if this limit exists. In the examples studied in this paper, I adopt conventions that give
\begin{align}
R(0) = \mathds{1}\ ,\qquad \lim_{u\to \infty}R_j(u)\propto B_j\ ,\qquad  \lim_{u\to -\infty}R_j(u)\propto \overline{B}_j .
\label{Rlimit}
\end{align}
For example, for the completely packed loop model the braid generators are
\begin{align}
B= q^{-\tfrac12}\mathord{\vcenter{\hbox{\includegraphics[scale=0.45]{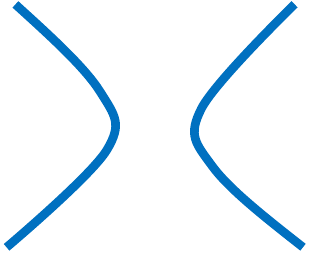}}}}\ -\  q^{\tfrac12} \mathord{\vcenter{\hbox{\includegraphics[scale=0.45]{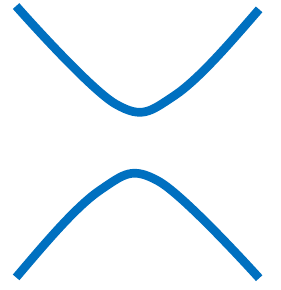}}}}\ ,\qquad\quad
\overline{B}= q^{\tfrac12} \mathord{\vcenter{\hbox{\includegraphics[scale=0.45]{loopv.pdf}}}}\ -\  q^{-\tfrac12}\mathord{\vcenter{\hbox{\includegraphics[scale=0.45]{looph.pdf}}}}\ .
\label{Bloop}
\end{align}
with a choice of overall phase. 

Neither the braid group nor a fusion category alone is sufficient to compute a knot invariant: the two must be combined into a {\em braided tensor category}. Useful reviews for physicists can be found in \cite{Moore1989,Kitaev2006,Bondersonthesis}. The additional data needed is in the relation 
\begin{align}
  \mathord{\vcenter{\hbox{\includegraphics[scale=0.45]{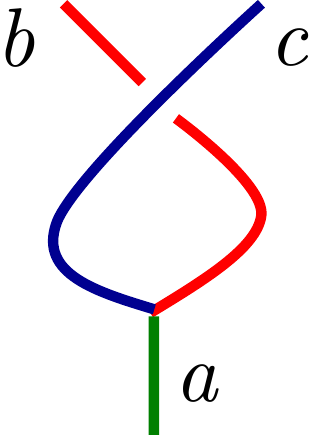}}}}\ =\  \Omega_a^{bc}\;
 \mathord{\vcenter{\hbox{\includegraphics[scale=0.5]{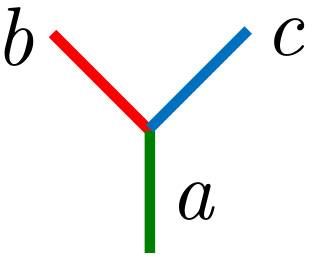}}}}\ ,\qquad\quad
   \mathord{\vcenter{\hbox{\includegraphics[scale=0.45]{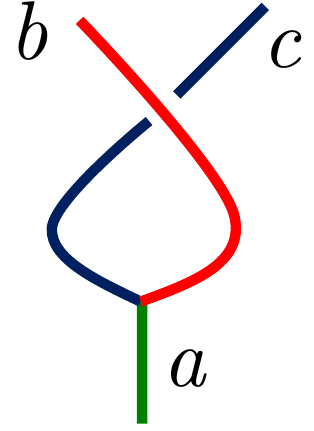}}}}\quad=\quad \Big(\Omega_a^{bc}\Big)^{-1}
   \ 
 \mathord{\vcenter{\hbox{\includegraphics[scale=0.5]{vertex.pdf}}}}\ .
\label{twist1}
\end{align}
The {\em twist factors} $\Omega_a^{bc}$ are roots of unity in a braided tensor category. They can be written in the form
\begin{align}
\Omega_a^{bc} = \nu_a^{bc}\, e^{i\pi (\Delta_b+\Delta_c-\Delta_a)}
\label{twistspin}
\end{align}
where the rational number $\Delta_a$ is called the {\em topological spin} of simple object $a$. The other coefficient $\nu_a^{bc}=\pm 1$ is an annoying sign, with the special case $\nu_0^{bb}$ known as the {\em Frobenius-Schur indicator}. 


No more data than that in \eqref{twist1} need be added to the fusion category data, as by using \eqref{joinstrands} the braid can be written as 
\begin{align}
  \mathord{\vcenter{\hbox{\includegraphics[scale=0.45]{overcrossing.pdf}}}}\quad=\ \sum_\chi \sqrt{\frac{d_\chi}{d_b d_c}}
 \mathord{\vcenter{\hbox{\includegraphics[scale=0.40]{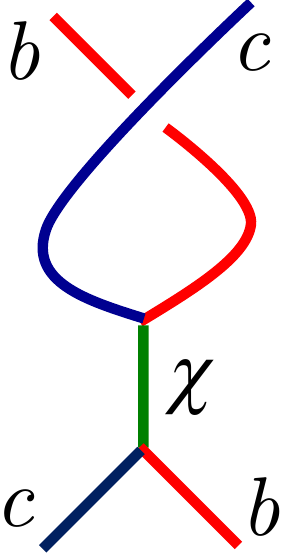}}}}\quad =\ \sum_\chi 
  \Omega_a^{bc}\sqrt{\frac{d_\chi}{d_b d_c}}
\  \mathord{\vcenter{\hbox{\includegraphics[scale=0.45]{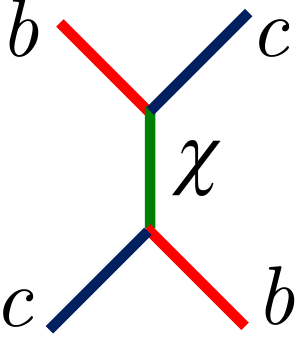}}}}\ .
\label{braidtwist}
\end{align}
The undercrossing is given by the same relation with all $\Omega_r\to \Omega_r^{-1}$. 
The expression  \eqref{braidtwist} of the braid as a sum over fusion channels is called a {\em skein relation}. When the braided tensor category is built from a quantum-group algebra, the coefficients are related to those of the universal $R$ matrix  \cite{Drinfeld1985,Reshetikhin1988,Kirillov1990,Khoroshkin1991}. The explicit matrix elements of braid generators acting on a fusion tree then can be computed explicitly by combining \eqref{braidtwist} with a calculation virtually identical to that leading to \eqref{Pmatrixelements}.

Combining braiding and fusing allows one to compute topological invariants for knots and links. Lines out of the plane of a fusion diagram can be deformed using the ``hexagon equation''
\begin{align}
  \mathord{\vcenter{\hbox{\includegraphics[scale=0.45]{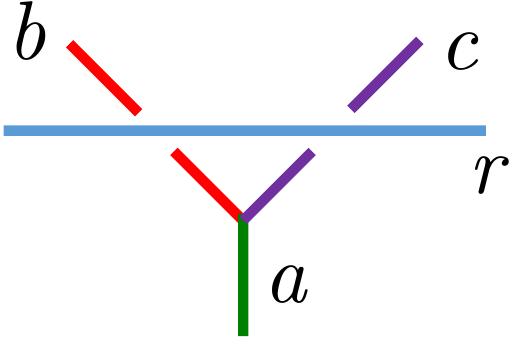}}}}\quad=\quad
 \mathord{\vcenter{\hbox{\includegraphics[scale=0.45]{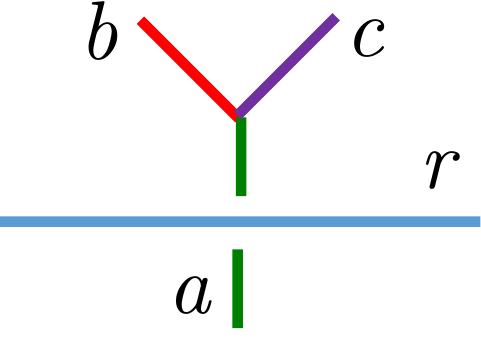}}}}\ .
\label{vertexbraid}
\end{align}
A useful identity for twist factors comes from setting $r=c$ and gluing together the corresponding lines in \eqref{vertexbraid}, and then undoing the resulting twists:
\begin{align}
\Omega_0^{cc} = \Omega_a^{bc}\, \Omega_{b}^{ac}\qquad\implies\quad \nu_0^{cc} = \nu_a^{bc} \nu_{b}^{ac}\ .
\label{nuid}
\end{align}
The latter relation comes from using \eqref{twistspin}.

Using \eqref{braidtwist} allows a knot or link to be reduced to a fusion diagram. The evaluation using the category gives a topological invariant, up to one subtlety. The first Reidemeister move is
\begin{align}
  \mathord{\vcenter{\hbox{\includegraphics[scale=0.45,angle=180]{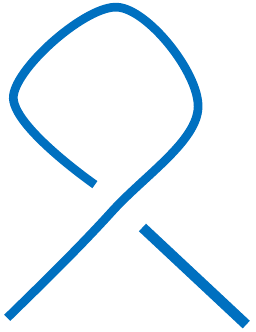}}}}\quad=\quad
 \mathord{\vcenter{\hbox{\includegraphics[scale=0.45,angle=180]{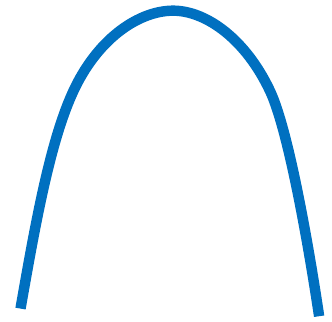}}}}\ .
\label{Reidemeister1}
\end{align}
Setting $c=b=\rho$ and $a=0$ in \eqref{twist1} yields this relation up to the phase, which therefore must be cancelled out in a topological invariant like the Jones polynomial. The easiest method is to ``frame'' the knot by treating it as a ribbon, so that \eqref{Reidemeister1} corresponds to a $2\pi$ twist of the ribbon (proof: try it with a belt). The (signed) number of these twists is called the writhe $w$, and each fusion diagram must be multiplied by the corresponding $(\Omega_0^{\rho\rho})^{-w}$ to obtain a topological invariant. While keeping track of the writhe is a pain in the calculation of a knot invariant, the phases in \eqref{twist1} are an essential part of constructing a conserved current.

The skein relations for completely packed loops \eqref{Bloop} are found from \eqref{braidtwist} by supplying the appropriate twist factors. 
One can use the category $su(2)_k$ arising from the quantum-group algebra $U_q(\mathfrak{sl}_2)$, which in turn is constructed from a deformation of the Lie algebra $\mathfrak{sl}_2$. The objects are labeled by their ``spin'' $0,\,\tfrac12,1,\dots,\tfrac{k}{2}$, and the fusion algebra is a truncated version of that of corresponding representations of $sl(2)$. It can be found in section 2.1 of \cite{Aasen2020}. For any $k>1$, setting $\rho=\tfrac12$ yields the fusion $\rho\otimes\rho = 0\oplus 1$ needed to construct loop models. The quantum dimensions are 
\begin{align}
d_s = \frac{q^{s+\frac12}- q^{-s-\frac12}}{q-q^{-1}}\ ,\qquad\quad q=e^{i\frac{\pi}{k+2}}
\label{dmin}
\end{align}
so in particular $d_{\frac12}=q+q^{-1}$. Here (see the discussion in section \ref{sec:examples})
\begin{align}
\Delta_a=\frac{a(a+1)}{k+2}\ ,\qquad \nu_a^{bc} = (-1)^{b+c-a}\quad\hbox{ for }su(2)_k\ .
\label{su2k}
\end{align}
Using  \eqref{braidtwist} and \eqref{squigglyremove} indeed yields \eqref{Bloop}. Another braided category $\mathcal{A}_{k+1}$ with the same objects and fusion rules as $su(2)_k$ arises from the $\Phi_{1,s}$ fields in the minimal models of conformal field theory \cite{Belavin84,Friedan1984}. The corresponding twist factors come from
\begin{align}
\Delta_a = a^2-\frac{a(a+1)}{k+2}\ ,\qquad\quad\nu_a^{bc}=1\quad \hbox{ for }\mathcal{A}_{k+1}\ ,
\label{Ak1}
\end{align}
giving \eqref{Bloop} with $q\leftrightarrow q^{-1}$ (or equivalently exchanging $B$ and $\overline{B}$).

\subsection{Defining the currents} 
\label{sec:currentdef}

I now turn to the central topic of this paper, defining and finding conserved non-local currents. 
The canonical example of such a current is the fermion operator in the critical Ising quantum spin chain.  It is non-local in the sense that when acting on the spin Hilbert space, the fermion operator at site $j$ flips all the spins with $j'<j$. This flipping is a very special sort of non-locality in that it commutes with all the Hamiltonian generators except those acting non-trivially at site $j$. An elementary computation shows that such a fermion operator obeys a conservation law. 

Such currents are very naturally and generally defined using braided tensor categories, both in geometric and height models. The partition function is written as an expansion over fusion diagrams as described in \eqref{Zgeom}, and expectation values of the current operators are computed by modifying the weights in each term of the sum. One very nice feature of this setup is this modification is done {only} to the fusion diagram, and hence only affects the topological part of each weight.

The current-current expectation value $\langle \overline{J}(w)\,J(z)\rangle$ is defined by modifying each fusion diagram to include another strand with label $\phi\ne 0$ and terminating it at two edges $w$ and $x$ of the square lattice. In the completely packed model, the simple object $\phi\ne 0$ must obey $\rho\in \phi\otimes\rho$ for the termination to be possible. The ensuing trivalent vertex for $J (z)$ is defined so that the $\phi$-strand is pointing upward or leftward, with $\overline{J}$ down or to the right: 
\begin{align}
J:\quad\mathord{\vcenter{\hbox{\includegraphics[scale=0.45]{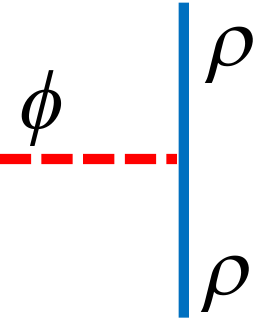}}}}\ ,\qquad
 \mathord{\vcenter{\hbox{\includegraphics[scale=0.45]{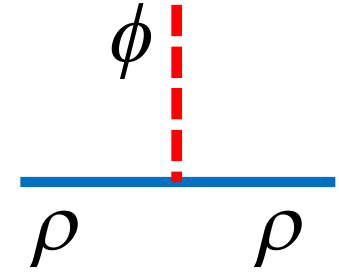}}}}\ ,\hspace{1in}
\overline{J}:\quad\mathord{\vcenter{\hbox{\includegraphics[scale=0.45]{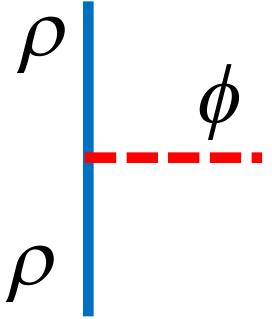}}}}\ ,\qquad
 \mathord{\vcenter{\hbox{\includegraphics[scale=0.45]{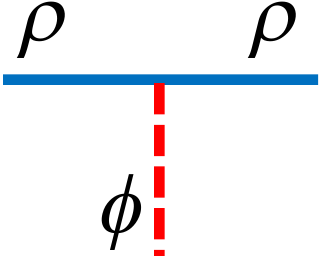}}}}\ .
\end{align}
The $\phi$ strand is drawn dashed solely as a visual aid -- it is treated as any other strand. Any time the $\phi$-strand meets a $\rho$ line, the ensuing intersection is defined to be an {\em over}crossing. The resulting diagram $\mathcal{F}_{w,z}$ involves both fusion and braiding. Then for any completely packed geometric model
\begin{align}
\big\langle \overline{J}(w)\,J(z)\big\rangle\equiv \frac{1}{Z_\rho}\sum_{\mcf_{w,z}} {\rm eval}_{\mathcal{C}}[{\mcf_{w,z}}]\, \prod_f   \frac{\sqrt{d_{\chi_f}}}{d_\rho}A_{\chi_f}\ .
\label{Jdef}
\end{align}
In a picture,
\begin{align}
\big\langle \overline{J}(w)\,J(z)\big\rangle\equiv \frac{1}{Z_\rho} \hbox{ eval}\  \mathord{\vcenter{\hbox{\includegraphics[scale=0.45]{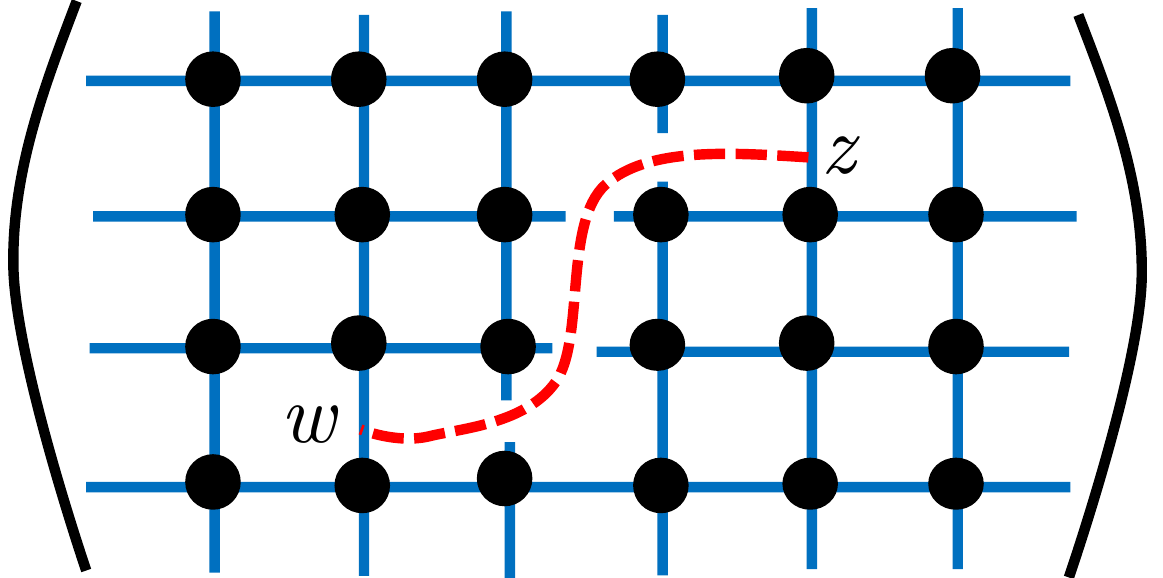}}}}\ .
\end{align}
using as always \eqref{crackopen} to obtain the sum over trivalent vertices.  The relation \eqref{bubbleremoval} requires that this correlator is non-vanishing only for $\mathcal{F}$ that connect $z$ and $w$. When $\rho$ is not simple, one may consider more complicated currents $J^{(\phi)}_{ab}(z)$ that change the label on an edge.

Because of \eqref{vertexbraid},  the path of the $\phi$-strand away from the vertices can be deformed without changing the evaluation. The currents therefore are very gently non-local, as the $\phi$-strand away from the vertices is a lattice {topological defect}. As discussed in depth in \cite{Aasen2016,Aasen2020}, topological defects can be found in any (not necessarily integrable or critical) two-dimensional classical lattice model built using a fusion category.  These defects can branch, and this structure can be utilised to define multi-point correlators of the currents. Although such expectation values remain independent of local deformations of the paths, they will depend on how the various $\phi$-strands pass over each other or fuse together. Understanding such multipoint correlators likely would be interesting, but I will discuss them no further here.

The currents are defined in height models simply by gluing the appropriate vertices to the corresponding $\rho$-legs of fusion tree, with the $\phi$-strand braiding appropriately. The action on the tree can be worked out using $F$ moves and bubble removal as done for the projectors in \eqref{Pmatrix}. 

\subsection{The conserved-current relation}

A braided tensor category provides a natural way not only to define currents, but find ones that are conserved. Current conservation in essence amounts to a vanishing lattice divergence at every vertex where the $\rho$ lines meet. In a picture,
\begin{align}
\mathord{\vcenter{\hbox{\includegraphics[scale=0.45]{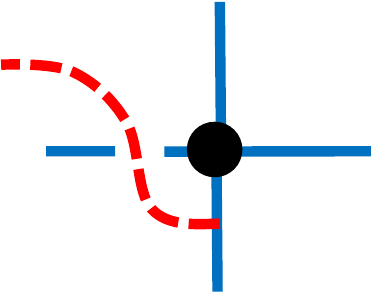}}}}\quad+\quad \mu\ 
\mathord{\vcenter{\hbox{\includegraphics[scale=0.45]{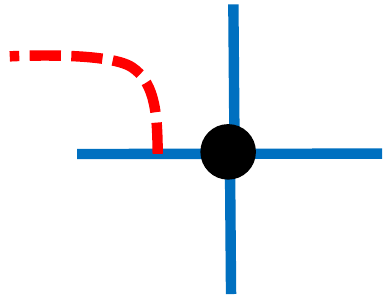}}}}\quad =\quad
\mathord{\vcenter{\hbox{\includegraphics[scale=0.45]{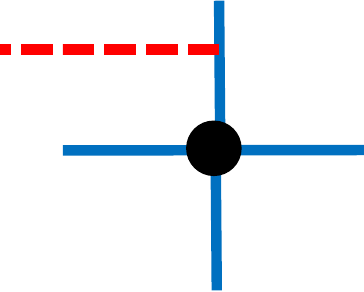}}}}\quad+\quad \mu\ 
\mathord{\vcenter{\hbox{\includegraphics[scale=0.45]{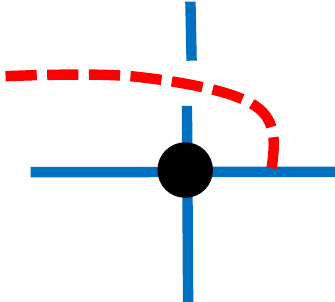}}}}\quad.
\end{align}
In an equation,
\begin{align}
J(x,y) + \mu J(x-1,y+1) = J(x,y+2) + \mu J(x+1,y+1)\ ,
\label{Jcons}
\end{align}
where $z$ is written in Cartesian coordinates $(x,y)$ and the lattice spacing is 2 so that edges have $x+y$ even. This operator equation means that the corresponding sum over two-point functions vanishes for any fixed $w$. The coefficient $\mu$ is a complex number, and can be thought of as a rescaling of the coordinates in the interpretation of \eqref{Jcons} as a vanishing divergence.
The corresponding charge is simply $Q=\sum_{n=1}^L J(x+2n,y)$, and is conserved (i.e.\ independent of $y$) up to boundary terms. 


The conservation law \eqref{Jcons} for fractional-spin currents in lattice models appeared long ago, going back at least to Bernard and Felder \cite{Bernard1991} in 1991.\footnote{I am grateful to Denis Bernard for this observation, made at my seminar on this work at the MSRI in 2012.} They showed how quantum-group algebras give a method for defining the vertices and finding solutions to the relation. Their work does not seem to have been widely noticed, perhaps because of the work involved in doing explicit calculations using the representation theory of quantum-group algebras.

The same relation was reintroduced in the interest of finding ``discretely holomorphic" operators \cite{Smirnov2006,Duminil2012}. An operator of spin $s$ in conformal field theory picks up a phase $e^{is\theta}$ under rotation by some angle $\theta$, just as the operator $J$ picks up $e^{i\pi \Delta_\phi}$ under twists by $\pi$. If the lattice model has a continuum limit, the spin of the CFT operator corresponding to $J$ should be $\pm\Delta_\phi\,$mod$\,1$, an observation confirmed numerically in many examples.  The idea of \cite{Smirnov2006} goes further, making the observation that if \eqref{Jcons} is written as $J_1 + \mu J_2 - J_3 -\mu J_4=0$ and $\mu$ is chosen appropriately, it amounts to a lattice analog of the vanishing of a contour integral, the lattice Cauchy-Riemann equation around one vertex. One then might hope that such currents  become holomorphic in the continuum limit, enabling a rigorous demonstration of conformal invariance emerging from a lattice model \cite{Smirnov2006,Duminil2012,Chelkak2012}. This approach works spectacularly for a few cases like Ising where the continuum field theory is free \cite{Chelkak2020}, but otherwise has not borne fruit. The reason presumably is that conserved-current condition (\ref{Jcons}) alone is insufficient to prove any form of holomorphicity, because it provides only half the constraints needed:  with twice as many edges as vertices on the square lattice, there are twice as many degrees of freedom as constraints. Indeed a counterexample occurs in the quantum three-state Potts chain. The lattice parafermion operator defined in \cite{Fradkin80} satisfies (\ref{Jcons}) \cite{Riva06} (as redone below in section \ref{sec:para}). However, in the scaling limit the lattice operator becomes the linear combination of two operators with the same $s\,$mod$\;1$, only one of which is holomorphic \cite{Mong2014}.

Nevertheless, one learns a great deal from \eqref{Jcons} by taking a different point of view.
A braided tensor category gives a natural definition of a current and the topological part of the Boltzmann weights, but it does not fix the amplitudes $A_\chi(u)$ determining the local part.
As only one Boltzmann weight appears in each term in \eqref{Jcons}, demanding current conservation gives a linear equation for these amplitudes. 
Solving it is a fairly straightforward exercise in loop models (completely packed and otherwise) and a few simple local models.  Cardy and others indeed found many more solutions by direct calculation \cite{Riva06,Rajabpour07,Cardy2009,Ikhlef2010,Ikhlef11,deGier2012,Alam2014,
Bondesan14,Ikhlef15,Chelkak2016,Ikhlef16,Chelkak2020}. More importantly, he made the very interesting observation that in all cases, the Boltzmann weights also satisfy the Yang-Baxter equation \eqref{YBE} \cite{Cardy2009}. As the latter is trilinear, much more brute force needed to solve it, and so utilising \eqref{Jcons} is much more efficient, not to mention elegant.

\section{Boltzmann weights from conserved currents}
\label{sec:central}

I show here how to use a braided tensor category to find a wide class of solutions to the conserved-current relation \eqref{Jcons}. The Boltzmann weights are given in terms of the category data and $\mu$, the latter turning into the spectral parameter. The method works for local and non-local models alike. In the completely packed case, the answer is rather simple, and in many special cases reduces to one obtained using quantum-group algebras. In these cases, as well as all others I have checked, the Boltzmann weights give a trigonometric solution of the Yang-Baxter equation. Solving the conserved-current relation therefore seems to give a linear method of Baxterisation.


Except for $\mu$, the quantities in the conserved-current relation are all defined via a braided tensor category. Manipulating the diagrams using $F$ moves and twisting does not change the evaluations,  and so leaves the correlators invariant. I explain here how such elementary manipulations can be used to put all four diagrams in \eqref{Jcons} in a common form, making finding solutions easy. 

The common form is found by writing each vertex as a sum over fusion channels using \eqref{crackopen}, and then using $F$ moves and twisting to move the $\phi$ strand into the centre to fuse with $\chi$.
A diagram requiring only an $F$ move is 
\begin{align}
\mathord{\vcenter{\hbox{\includegraphics[scale=0.45]{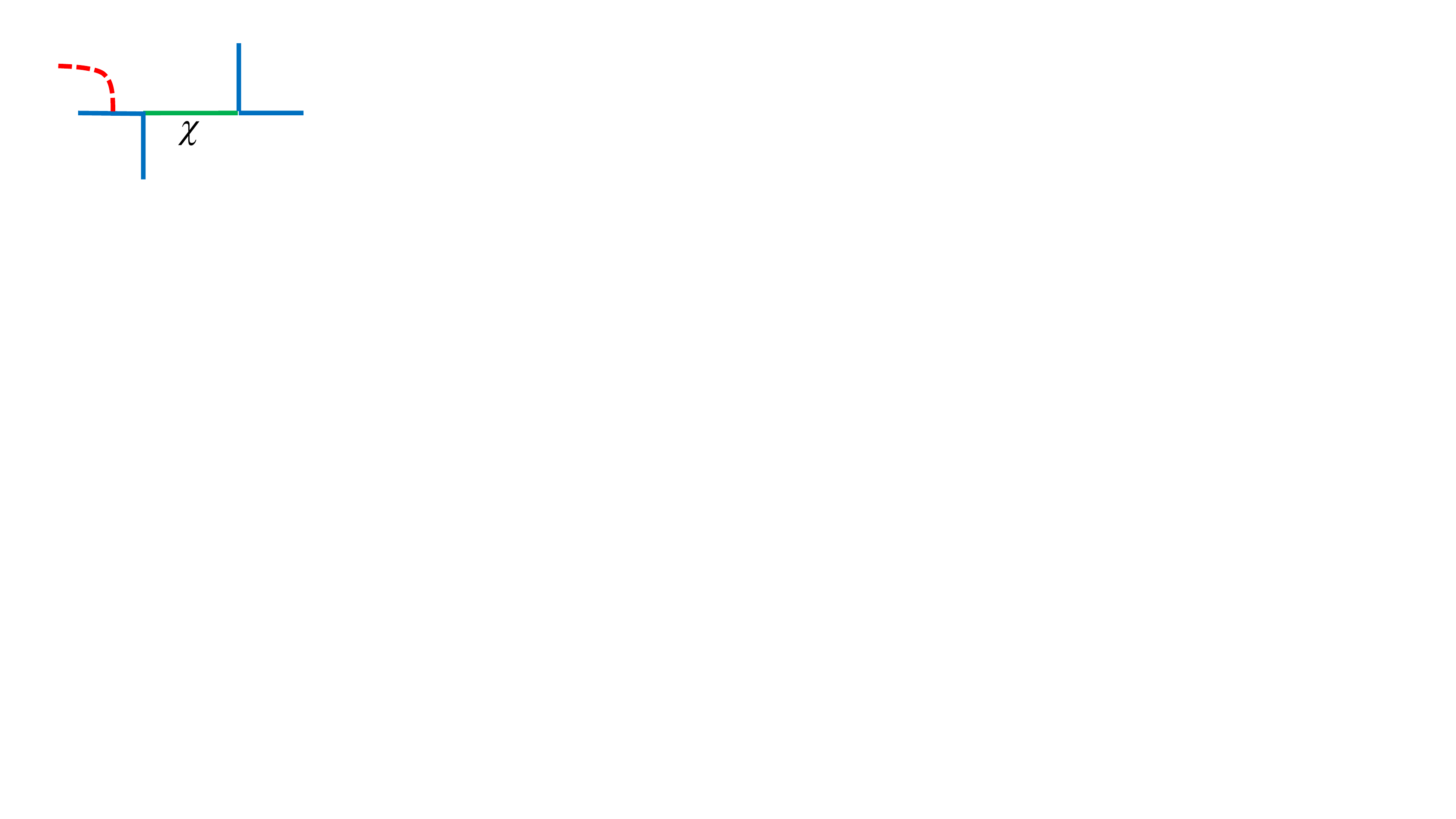}}}}\quad=\ 
\sum_{a} F_{\rho a} \begin{bmatrix} \phi&\chi\\ \rho&\rho\end{bmatrix}\ 
\mathord{\vcenter{\hbox{\includegraphics[scale=0.45]{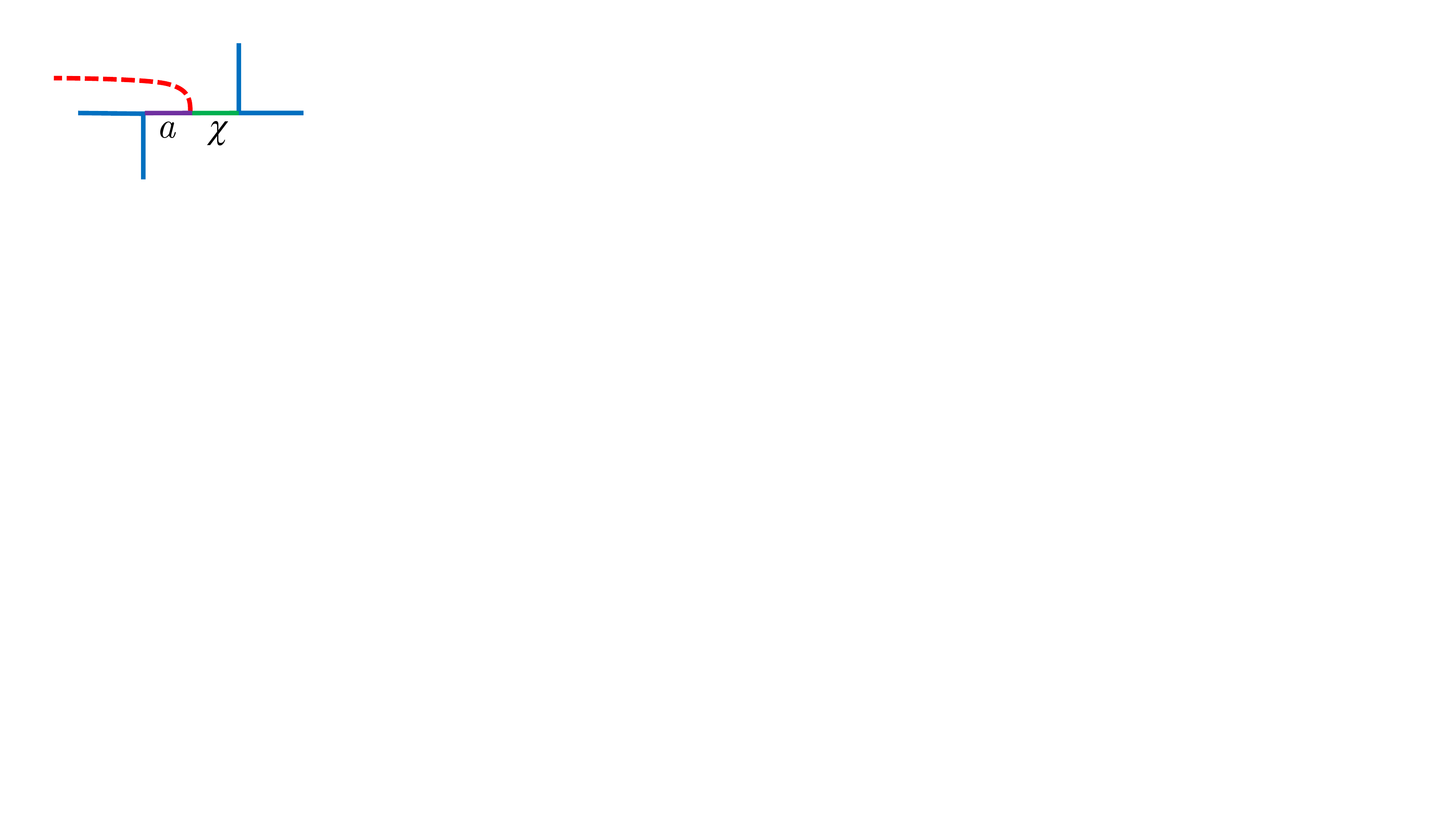}}}}\ ,
\end{align}
where as always the dashed line is labelled $\phi$ and unlabelled solid lines $\rho$.
Another simple case is 
\begin{align}
\mathord{\vcenter{\hbox{\includegraphics[scale=0.45]{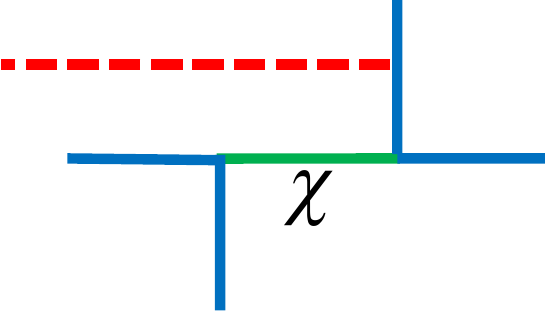}}}}\quad=\ 
\sum_{b} F_{\rho b} \begin{bmatrix}  \chi&\phi\\ \rho&\rho\end{bmatrix}\ 
\mathord{\vcenter{\hbox{\includegraphics[scale=0.45]{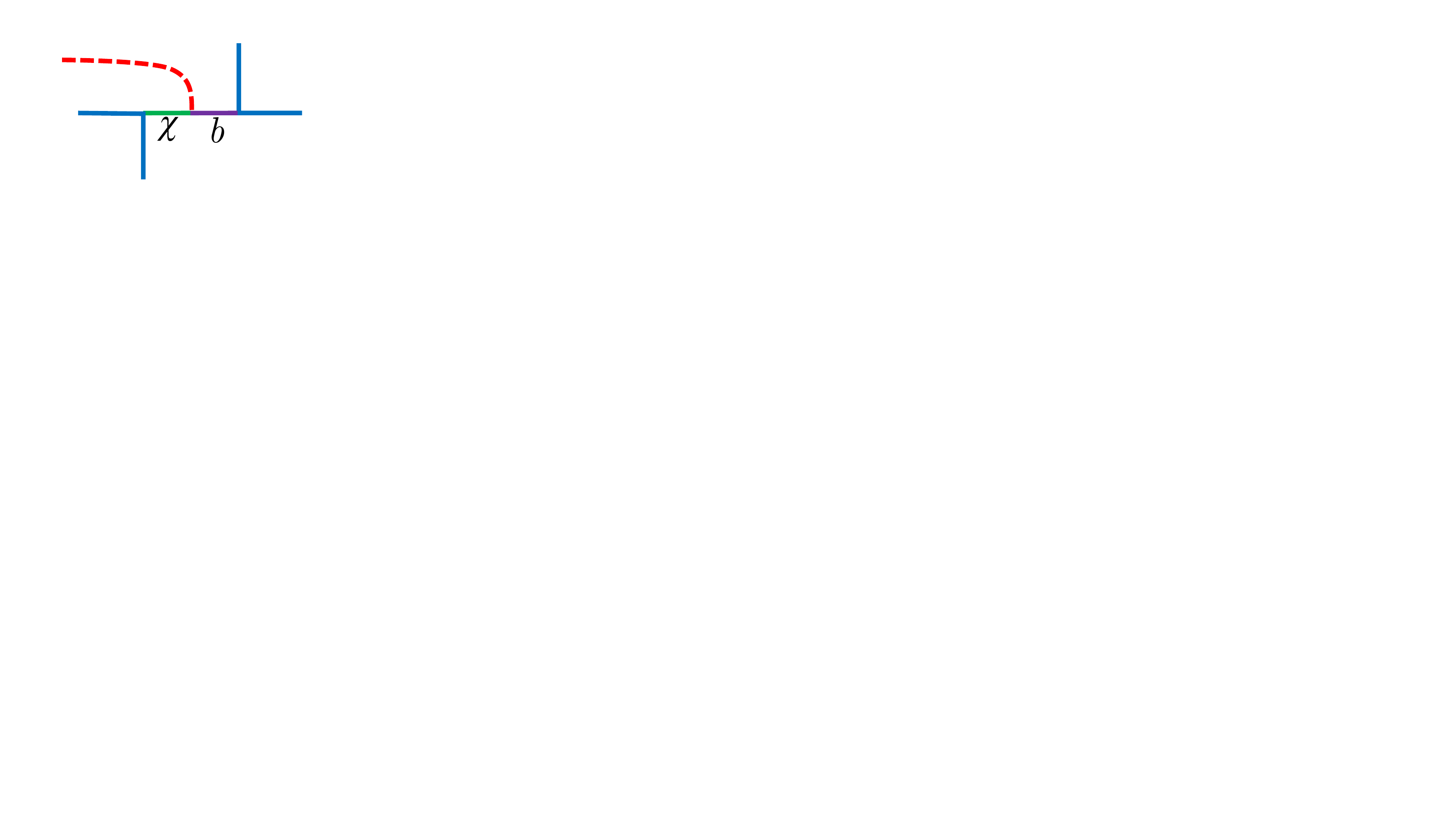}}}}\ .
\end{align}
The other two require several twists, yielding
\begin{align}
\begin{split}
\mathord{\vcenter{\hbox{\includegraphics[scale=0.45]{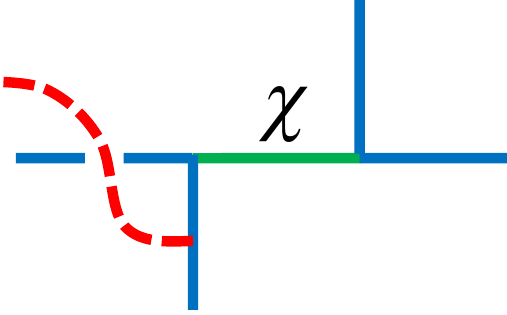}}}}\ 
&=\big(\Omega_{\rho}^{\rho\chi}\big)^{-1}\; \mathord{\vcenter{\hbox{\includegraphics[scale=0.5]{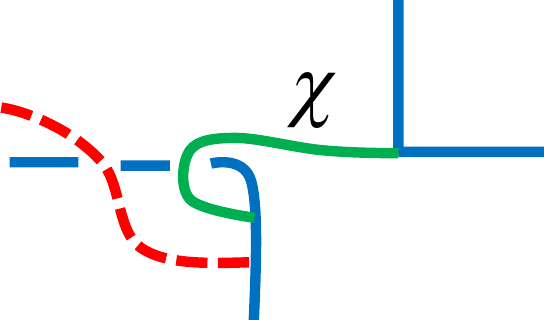}}}}\ =
\big(\Omega_{\rho}^{\rho\chi}\big)^{-1}\; \sum_{a} F_{\rho a} \begin{bmatrix}  \phi&\chi\\ \rho&\rho\end{bmatrix}\ \mathord{\vcenter{\hbox{\includegraphics[scale=0.5]{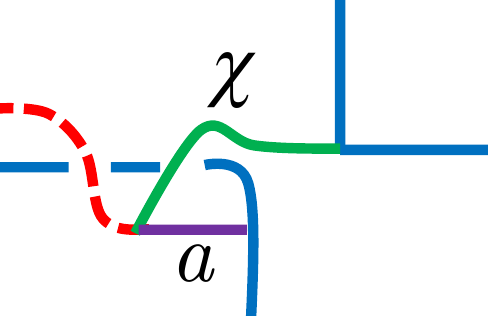}}}}\cr
&=
\big(\Omega_{\rho}^{\rho\chi}\big)^{-1}\; \sum_{a} F_{\rho a} \begin{bmatrix}  \phi&\chi\\ \rho&\rho\end{bmatrix}\ \mathord{\vcenter{\hbox{\includegraphics[scale=0.5]{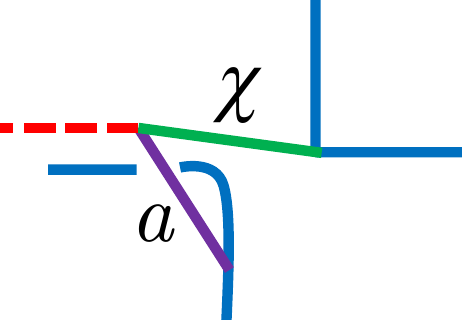}}}}\cr
&=
\big(\Omega_{\rho}^{\rho\chi}\big)^{-1}\; \sum_{a} \Omega_{\rho}^{\rho a} F_{\rho a} \begin{bmatrix}  \phi&\chi\\ \rho&\rho\end{bmatrix}\ \mathord{\vcenter{\hbox{\includegraphics[scale=0.45]{solve2-2.pdf}}}}
\end{split}
\end{align}
and via a similar sequence of moves
\begin{align}
\mathord{\vcenter{\hbox{\includegraphics[scale=0.5]{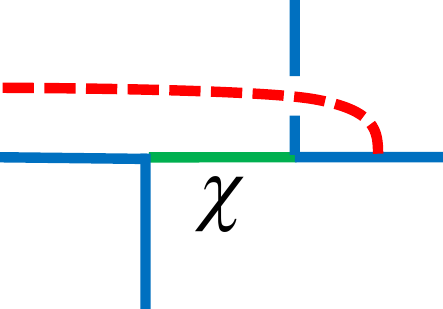}}}}\ 
= \Omega_{\rho}^{\rho \chi}\;\sum_{b} \big(\Omega_{\rho}^{\rho b}\big)^{-1}\, F_{\rho b} \begin{bmatrix}  \chi&\phi\\ \rho&\rho\end{bmatrix}\ \mathord{\vcenter{\hbox{\includegraphics[scale=0.5]{solve1-2.pdf}}}}\ .
\end{align}

These four relations put all the terms in \eqref{Jcons} into a common form, with two lines fusing in the middle to give $\phi$. Summing over fusion channels from \eqref{crackopen} as well as those coming from the $F$ move means that for each such $a,b$ obeying $N_{a\phi}^b=N_{b\phi}^a\ne 0$, 
\begin{align}
A_b\,\sqrt{d_b} \;F_{\rho a} \begin{bmatrix} \phi&b\\ \rho&\rho\end{bmatrix}\Big(\Omega_{\rho}^{\rho a}\big(\Omega_{\rho}^{\rho b}\big)^{-1}+\;\mu\Big)
=A_a\,\sqrt{d_a} \;F_{\rho b} \begin{bmatrix} a&\phi\\ \rho&\rho\end{bmatrix}\Big(\;1+\;\mu\,\Omega_{\rho}^{\rho a}\big(\Omega_{\rho}^{\rho b}\big)^{-1}\Big)\ .
\end{align}
Happily, the identity \eqref{usefulid} means the $F$ symbols and quantum dimensions cancel, leaving 
\begin{align}
\boxed{\quad
\frac{A_b}{A_a}\  =\ \frac{\Omega_{\rho}^{\rho b} +\;\mu\,\Omega_{\rho}^{\rho a}}
{\Omega_{\rho}^{\rho a} +\;\mu\,\Omega_{\rho}^{\rho b}}\qquad\quad\hbox{when }\ N_{a\phi}^b\ne 0\,.\quad}
\label{central}
\end{align}
when as always $a,b\in \rho\otimes\rho$, and $\rho \in \phi\otimes\rho$. When all objects are self-dual, the latter condition implies $\phi\in\rho\otimes\rho$ as well.

The elegant relation \eqref{central} arose long ago in models defined using quantum-group algebras and with $\phi$ a particular representation (the adjoint in the untwisted case)  \cite{Zhang1990,Delius1994,Delius1995}.  Here it gives solutions of Jimbo's equation \cite{Jimbo1986} for Baxterising a representation of a quantum-group algebra. The relations (\ref{Jcons},\ref{central}) use only the data coming from the braided tensor category, so the approach here provides not only a generalization, but a shortcut to the result. The advantage of the quantum-group approach, however, is that weights satisfying Jimbo's equation satisfy the Yang-Baxter equation, whereas I have only proved current conservation \eqref{Jcons}. Nevertheless, it is natural to expect \eqref{central} implies \eqref{YBE} well outside quantum-group algebras, as requiring current conservation already highly constrains the theory. Indeed, an analogous argument in Lorentz-invariant field theory states that having even one additional conservation law not commuting with the Poincare algebra is sufficient to make the model integrable \cite{Shankar77,Dorey97}. 

\section{Boltzmann weights with conserved currents}
\label{sec:examples}

The constraint \eqref{central} is the central result of this paper. Boltzmann weights satisfying it admit a conserved current defined via (\ref{Jdef}\,,\ref{Jcons}) in the completely packed model ($\rho$ simple). 
The simplicity of \eqref{central} is rather striking. It depends only on the twist factors and the fusion algebra, with the current type $\phi$ only entering the latter. 
No conditions are placed on $\mu$. 

Here I describe how to find categories and objects for which there is a solution. I find many examples where it works and explain how to see when it does not.
Although most of the weights found here are known to satisfy the Yang-Baxter equation, one series seems not to have arisen previously. Moreover, only a few of the conserved currents derived in this section have been analysed before.

\subsection{A few general considerations}

\paragraph{Explicit formulas for category data.} The twist-factor ratio needed can be rewritten as
\begin{align}
\frac{\Omega_{\rho}^{\rho b}}{\Omega_{\rho}^{\rho a}}\ =\ \frac{\Omega_{a}^{\rho\rho}}{\Omega_{b}^{\rho \rho}}\ =\ 
\nu_{a}^{\rho\rho} \nu_b^{\rho\rho} e^{i\pi(\Delta_b-\Delta_a)} \ ,
\end{align} 
where the first equality comes from using \eqref{nuid} and the second from \eqref{twistspin}. An explicit expression for $\Omega_{\rho\rho}^b$ in any modular tensor category can be found in proposition 2.3 in \cite{Bonderson2018}. A modular tensor category extends the braided tensor category to allow for fusion diagrams on surfaces, including data for modular transformations. The explicit expression involves these data, the modular $S$ matrix.
Simpler expressions exist for any category $g_k$ built from a quantum-group algebra $U_q(\mathfrak{g})$ \cite{Gomez1996}. The topological spin is proportional to the quadratic Casimir $C_\mathfrak{g}$ of the corresponding representation of the (undeformed) Lie algebra $\mathfrak{g}$:
\begin{align}
\Delta_a = \frac{C_\mathfrak{g}(a)}{k + h_\mathfrak{g}}\qquad\ \hbox{ for } g_k\ .
\label{Deltadef}
\end{align}
The level $k$ is a positive integer, while $h_{\mathfrak{g}}$ is the dual Coexter number of $\mathfrak{g}$ (the quadratic Casimir of the adjoint representation). The sign $\nu_a^{bc}=\pm 1$ is determined by the symmetry (+1) or antisymmetry ($-1$) of the invariant tensor coupling the representations $b$ and $c$ into $a$. It is worth noting that much useful information about simple Lie algebras such as tensor products and quadratic Casimirs may be accessed using the Mathematica package LieART \cite{Feger2020}.

\paragraph{Tensor-product graphs}
A solution of \eqref{central} does not automatically exist for a given $\rho\in\phi\otimes\rho$. For $r$ objects in $\rho\otimes\rho$ and $0\in\rho\otimes\rho$, there are up to $r(r-1)/2$ possible distinct equations in \eqref{central} but only $r-1$ independent amplitude ratios. Of course, some $N_{a\phi}^b=0$ may be zero, information that can be summarised  conveniently in a {\em tensor product graph} \cite{Zhang1990}. The vertices of this graph are labeled by the objects in $\rho\otimes\rho$, and two vertices $a,b$ share an edge if $N_{a\phi}^b\ne 0$, so that each edge corresponds to one relation. For example, when $\rho\otimes\rho = 0 \oplus 1$ as for completely packed loops, the only possible non-trivial label for the current is $\phi=1$, and the tensor-product graph is simply
$\ 0\; \raisebox{3pt}{\rule{0.5cm}{0.5pt}}\;1$.  When all the objects are self-dual as assumed above, $N_{a\phi}^b=N_{b\phi}^a$, so the edges of the graph do not need to be oriented. I give an example below in section \ref{sec:para} where this assumption is relaxed.

\subsection{Tree tensor-product graphs}
\label{sec:tree}

Euler's relation ensures that the number of edges of a tree (a graph with no cycles) is one less than the number of vertices. Thus when the tensor-product graph is a tree, the number of constraints coming from \eqref{central} is the same as the number of amplitude ratios. Requiring a conserved current exist then fixes all the Boltzmann weights. Here I discuss a variety of such examples, and find what may be a previously unknown (or at least undisplayed) solution of the Yang-Baxter equation.

\paragraph{Loops and ABF.}   The data for two braided tensor categories with the fusion rule  $\rho\otimes\rho = 0 \oplus 1$ are given in \eqref{su2k} \eqref{Ak1}. Using  \eqref{twistspin} to get the twist factors gives immediately
\begin{align}
\frac{A_0}{A_1}\ =
 \frac{1\;-\;\mu\,q^2}{\mu\;-\; q^2}
\quad\hbox{ for }su(2)_k\ ,\qquad\quad
\frac{A_0}{A_1}\ =\frac{1\;-\;\mu\,q^{-2}}{\mu\;-\; q^{-2}}
\quad\hbox{ for }\mathcal{A}_{k+1}\ .
\label{Aloop}
\end{align}
These amplitudes result in conserved currents in the completely packed loop models, recovering the results of \cite{Riva06}. The shadow-world construction described in section \ref{sec:heights} extends the result to the corresponding height models of Andrews, Baxter and Forrester \cite{Andrews1984}.  Even more exciting is the fact that weights satisfy the Yang-Baxter equation, as apparent from comparison with \eqref{YBEloop} with $\mu = e^u$ for $su(2)_k$ and $\mu=e^{-u}$ for $\mathcal{A}_{k+1}$.

\paragraph{One category, two solutions}
The next-simplest case is when $\rho\otimes\rho$ is the sum of three objects. The vector representation $V$ of  $so(n)_k$ or $sp(2m)_k$ with $n>2$, $m>1$ and $k\ge 2$ has
\begin{align}
{V}\otimes {V} = 0 \oplus \mathscr{A} \oplus {S}
\end{align} with $\mathscr{A}$ and $S$ the antisymmetric and symmetric representations respectively. In these categories $F$ moves do not allow trivalent vertices to removed as in \eqref{squigglyremove} in the loop model, and the corresponding Birman-Murakami-Wenzl algebra \cite{Birman1989,Murakami1990} generalises Temperley-Lieb to include intersections. 
Either $\mathscr{A}$ or $S$ can be used for $\phi$, with tensor product graphs
\begin{align}
0\; \raisebox{3pt}{\rule{0.5cm}{0.5pt}}\; \mathscr{A}\;\raisebox{3pt}{\rule{0.5cm}{0.5pt}}\;S \quad\hbox{ for }\ \phi=\mathscr{A}\ ,\qquad\quad
0\; \raisebox{3pt}{\rule{0.5cm}{0.5pt}}\; S\;\raisebox{3pt}{\rule{0.5cm}{0.5pt}}\;\mathscr{A} \quad\hbox{ for }\ \phi=S\ .
\end{align}
The signs needed to compute the Boltzmann weights are $\nu_{VV}^\mathscr{A}=-1$ are $\nu_{VV}^S=1$, as the names of the objects indicate. For $so(n)$, $\nu_{VV}^0=1$ and the quadratic Casimirs are $C_n(\mathscr{A})=h_{so(n)}=n-2 $ and $C_n(S)=n$. The ratios from \eqref{central} are then
\begin{align}
\begin{split}
&\phi=\mathscr{A}:\quad\ \ \frac{A_0}{A_\mathscr{A}}\ =
 \frac{1\;-\;\mu\,q^{n-2}}{\mu\;-\; q^{n-2}}\ ,\quad
\frac{A_S}{A_\mathscr{A}}\ =\frac{1\;-\;\mu\,q^{-2}}{\mu\;-\; q^{-2}}
\quad
\\[6pt]
&\phi=S:\qquad \frac{A_0}{A_S}\ =
 \frac{1\;+\;\mu\,q^{n}}{\mu\;+\; q^{n}}\ ,\qquad
\frac{A_\mathscr{A}}{A_S}\ =\frac{1\;-\;\mu\,q^{2}}{\mu\;-\; q^{2}}
\end{split}
\quad
\hbox{ for } so(n)_k\ \hbox{with } q=e^{i\frac{\pi}{n+k-2}}\ .
\label{Asonk}
\end{align}
For $sp(2m)$, $\nu_{VV}^0=-1$ along with $C_m(\mathscr{A}) =m$ and $C_m(S)=h_{sp(2m)}=m+1$, so 
\begin{align}
\begin{split}
&\phi=\mathscr{A}:\quad\ \ 
\frac{A_0}{A_\mathscr{A}}\ =
 \frac{1\;+\;\mu\,q^{m}}{\mu\;+\; q^{m}}\ ,
\qquad
\frac{A_S}{A_\mathscr{A}}\ =\frac{1\;-\;\mu\,q^{-1}}{\mu\;-\; q^{-1}}
\\[6pt]
&\phi=S:\qquad\frac{A_0}{A_S}\ =
 \frac{1\;-\;\mu\,q^{m+1}}{\mu\;-\; q^{m+1}}\ ,\qquad
\frac{A_\mathscr{A}}{A_S}\ =\frac{1\;-\;\mu\,q^{}}{\mu\;-\; q^{}}
\end{split}
\qquad\hbox{ for }sp(2m)_k\ \hbox{with } q=e^{i\frac{\pi}{m+k+1}} \ .
\label{Aspnk}
\end{align}

\paragraph{A new solution of Yang-Baxter?}
A nice aspect of the approach here is that distinct solutions for a given model stem arise naturally from different choices of $\phi$. In the quantum-group approach, the solutions come from very different places. The weights in (\ref{Asonk},\,\ref{Aspnk}) where $\phi$ is the adjoint representation ($\mathscr{A}$ for $so(n)$ and $S$ for $sp(2m)$) correspond to long-known solutions of the Yang-Baxter equation, found by rewriting the height-model weights of \cite{Jimbo1988} in terms of the projectors $P^{(0)}$, $P^{(\mathcal{A})}$ and $P^{(S)}$ defined in \eqref{Pmatrixelements}. These are the solutions of Jimbo's equation corresponding to untwisted Kac-Moody algebras \cite{Zhang1990,Delius1994}. The weights for $sp(2m)_k$ for $\phi=\mathscr{A}$ and $so(2l+1)_k$ for $\phi=S$ correspond to the solutions of Jimbo's equation for the twisted Kac-Moody algebras $A^{(2)}_{2m-1}$ and $A^{(2)}_{2l}$ respectively \cite{Kuniba1991,Delius1995}. (The case $l=1$ coming from $so(3)_k$ is known as the Izergin-Korepin $R$-matrix \cite{Izergin1980}.) The solution for $so(2l)_k$ with $\phi=S$ from \eqref{Asonk} seems to have appeared only implicitly in e.g.\ \cite{Fendley2001}.  Given how naturally it fits in with the others, I conjecture it also satisfies the Yang-Baxter equation.

\paragraph{Higher spins}
A solution for $\phi=1$ in the $su(2)_k$ or $\mathcal{A}_{k+1}$ categories exists for any $\rho$,  generalizing the $\rho=\tfrac12$ and $1$ results. The tensor-product graph is
\begin{align}
0\; \raisebox{3pt}{\rule{0.5cm}{0.5pt}}\; 1\;\raisebox{3pt}{\rule{0.5cm}{0.5pt}}\;2 \;\raisebox{3pt}{\rule{0.35cm}{0.5pt}}\ 
\dots\ \raisebox{3pt}{\rule{0.35cm}{0.5pt}}\; \hbox{min}(2s,\,k-2s)\ .
\label{tghigherspin}
\end{align}
The Boltzmann weights admitting a conserved current are found easily using \eqref{central} with the category data from \eqref{su2k} or \eqref{Ak1}, giving for $a=0,1\dots, \hbox{min}(2s,\,k-2s)$
\begin{align}
\frac{A_{a+1}}{A_a} = \frac{1 - \mu q^{-2(a+1)}}{\mu - q^{-2(a+1)}} \qquad\hbox{ for }su(2)_k\ \hbox{with } q=e^{i\frac{\pi}{k+2}} 
\end{align}
with $q\to q^{-1}$ for $\mathcal{A}_{k+1}$. 
These weights solve Jimbo's equation and hence Yang-Baxter, and were found long ago \cite{Date1987} using the fusion procedure \cite{Kulish1981}. Using a higher value of $\phi$ however leads to a more complicated tensor-product graph without a solution, as I discuss below.

\paragraph{Another two-solution case}
Taking $\rho$ to be a spinor representation in $so(n)_k$ leads to another tensor-product graph where all the objects lie in a line. Taking $\phi=\mathscr{A}$, the adjoint representation, gives solutions of Yang-Baxter as expected \cite{Delius1994}. An interesting feature for odd $n$ (where there is only one spinor representation) is that taking $\phi=V$ also leads to a tensor-product graph where all objects lie in a line. As explained in \cite{Delius1995}, this second solution corresponds to the twisted algebra $D^{(2)}_{(n+1)/2}$.

\paragraph{Exceptional cases}

Conserved currents can also be constructed for categories built on exceptional Lie algebras. One nice example comes from the $(G_2)_k$ category by taking $\rho$ to be the 7-dimensional vector representation (treating $G_2$ as a subalgebra of $so(7)$ \cite{Macfarlane2001}).  Its fusion is $V\otimes V=0\oplus V \oplus \mathscr{A} \oplus S$. Taking $\phi$ in the adjoint representation gives the tensor-product graph
\begin{align}
0\; \raisebox{3pt}{\rule{0.5cm}{0.5pt}}\;\mathscr{A}\;\raisebox{3pt}{\rule{0.5cm}{0.5pt}}\;S\;\raisebox{3pt}{\rule{0.5cm}{0.5pt}}\;V \qquad\hbox{ for }\ \phi=\mathscr{A}\ .
\end{align}
The quadratic Casimirs are $C_{G_2}(V)= 2$, $C_{G_2}(\mathscr{A})=4$, $C_{G_2}(S)=14/3$, while the signs are as in $so(7)$, namely $\nu_{VV}^0=\nu_{VV}^S=1$, $\nu_{VV}^V=\nu_{VV}^\mathscr{A}=-1$. Then \eqref{central} gives
\begin{align}
\frac{A_0}{A_\mathscr{A}}\ =
 \frac{1\;-\;\mu\,q^{4}}{\mu\;-\; q^{4}}\ ,\quad
 \frac{A_S}{A_\mathscr{A}}\ =
 \frac{1\;-\;\mu\,q^{-2/3}}{\mu\;-\; q^{-2/3}}\ ,\quad
\frac{A_V}{A_S}\ =\frac{1\;-\;\mu\,q^{8/3}}{\mu\;-\; q^{8/3}}\qquad \hbox{with } q=e^{i\frac{\pi}{k+4}} 
\end{align}
in agreement with \cite{Kuniba1991}. A similar calculation gives solutions for the fundamental representations of 
$E_7$ and $F_4$ \cite{Kim1990}.

\subsection{Tensor-product graphs with cycles}
\label{sec:para}

When a tensor-product graph contains a cycle, the relations \eqref{central} overconstrain the Boltzmann weights. Generically, there is no such solution for a given $\phi$. Moreover, for some $\rho$ there exists no $\phi$ yielding a conserved current, and so presumably no solution of the Yang-Baxter equation. However, sometimes the extra constraints can be satisfied, and I discuss a few examples here.

\paragraph{Escaping quantum groups with parafermions}

All the lattice models analysed in section \ref{sec:tree} can be built from a quantum-group algebra. Here I discuss an example that cannot, the integrable $\mathbb{Z}_M$-invariant clock models  \cite{Fateev82}. These models are built from the $\mathbb{Z}_M$ Tambara-Yamagami category \cite{Tambara1998}. It has $M+1$ objects, labeled $X$ and $a=0,1\dots M-1$, with fusion algebra
\begin{align}
a\otimes a' = (a+a')\,\hbox{mod}\,M\ ,\qquad X\otimes a = a\ ,\qquad X\otimes X = \sum_{a=0}^{M-1} a\  .
\end{align}
Taking $\rho=X$ then gives a height model where half the heights are $X$, with the other half any value $0,1,\dots M-1$.  The Boltzmann weights thus can be written in terms of the projectors $P^{(a)}$, which are given in category language in \cite{Aasen2020}. The topological spins are $h_a = a(M-a)/M$, the dimensions of the parafermion fields in the corresponding conformal field theory \cite{Zamolodchikov85}, while all $\nu=1$. 

The current $J$ defined by taking $\phi=1$ is known as the parafermion operator, found by generalising the Jordan-Wigner transformation \cite{Fradkin80}.  The fusion coefficients needed to define the tensor-product graph are $N_{a1}^{a+1} = 1$ (with indices interpreted mod $M$) and zero otherwise. Because the objects $a=1,\dots, M-1$ are not self-dual ($\overline{a} =M-a$), $N_{a\phi}^b\ne N_{b\phi}^a$ and the edges of the tensor-product graph need orientation. Putting an arrow pointing from $a\to a+1$ results in an $M$-sided oriented polygon, e.g.\ 
\setlength{\unitlength}{0.2cm}
\begin{center}\begin{picture}(9.8,8)
\put(0,4){$0$}\put(1.2,5.5){\vector(3,4){2}}\put(3.3,8.5){$1$}\put(4.5,9){\vector(1,0){4}}
\put(8.8,8.5){$2$}\put(10.2,8.4){\vector(3,-4){2}}\put(12.2,4){$3$}
\put(12.2,3.3){\vector(-3,-4){2}} \put(8.8,-0.5){$4$}\put(8.5,0){\vector(-1,0){4}}\put(3.3,-0.5){$5$}
\put(3.1,1){\vector(-3,4){2}}
\end{picture}
\end{center}
for $M=6$. For $J$ to be conserved \cite{Rajabpour07}, \eqref{central} relates the amplitudes as
\begin{align}
\frac{A_{a+1}}{A_a} = \frac{1 - \mu \omega^{a+\frac12}}{\mu - \omega^{a+\frac12}}\qquad\hbox{ with } \omega=e^{i\frac{2\pi}{M}} \ .
\label{Apara}
\end{align}
where $A_M\equiv A_0$. The fact that the tensor-product graph is a cycle means that there is one more equation in \eqref{central} than there are amplitude ratios, and so one consistency condition. It amounts to checking that \eqref{Apara} for both $a=0$ and $a=M-1$ indeed give the same ratio $A_{1}/A_0$. The ratios \eqref{Apara} are precisely those found in \cite{Fateev82} to satisfy the Yang-Baxter equation, with the identification $\mu=e^{i\alpha/M}\omega^{-1}$. Thus again demanding a conserved current results in an integrable model.

\paragraph{Other solutions with cycles}
Despite the results for parafermions,  a $\phi$ whose tensor-product graph has a cycle generally does not yield a conserved current. For example, consider $su(2)_k$ or  $\mathcal{A}_{k+1}$, with $\rho=3/2$, so that 
$\rho\otimes\rho = 0\oplus 1 \oplus 2 \oplus 3$ when $k\ge 6$. The tensor-product graph for $\phi=1$ given in \eqref{tghigherspin} is a tree, leading to a conserved current. However, taking $\phi=2$ gives a graph with a cycle:
\begin{center}\begin{picture}(10,2)
\put(0.7,-0.4){$0$}\put(1.9,0.2){\line(1,0){3.5}}\put(5.7,-0.4){$2$}\put(7,0.7){\line(2,1){2.5}}\put(9.7,1.7){1}
\put(10.1,1.3){\line(0,-1){1.8}}\put(9.7,-2){3}\put(7,0){\line(2,-1){2.5}}
\end{picture}
\end{center}
Using the appropriate data shows the only consistent solution to \eqref{central} is if $1=6\,{\rm mod}\,(k+2)$, which cannot be satisfied with $k\ge 6$. Changing $\rho$ typically makes matters worse, as the more objects in $\rho\otimes\rho$, the more difficult it is to avoid cycles in the tensor-product tree and the resulting constraints. 

Nevertheless, solutions of \eqref{central} still exist in a few cases with cycles. A number of examples where $\phi$ is the adjoint representation of the quantum-group algebra are described in \cite{Zhang1990,Delius1994}. Here I briefly describe one example not covered there, and apparently discussed in the literature only in the $k\to\infty$ rational limit \cite{Babichenko2002}. The example is $sp(2m)_k$ taking $\rho=S$ (the adjoint) and $\phi=\mathscr{A}$. The ensuing tensor-product graph is
\begin{center}\begin{picture}(19,7)
\put(0.7,3){$0$}\put(1.9,3.5){\line(1,0){3.5}}\put(5.5,3){$\mathscr{A}$}\put(7.6,4.7){\line(1,1){2}}\put(10,6.4){$S$}
\put(9.6,-0.8){\small $2\mu_2$}\put(7.4,2.7){\line(1,-1){2}}
\put(11.7,6.6){\line(1,-1){2}}\put(11,3){\small $2\mu_1$+$\mu_2$} \put(11.7,0.4){\line(1,1){2}} \put(17.5,3.5){\line(1,0){3}}
\put(21,3){\small $4\mu_1$}
\end{picture}
\end{center}
where the other representations are labelled by their highest weights in the standard conventions \cite{Feger2020}. 
The constraint coming from the cycle is satisfied because $\Omega^{SS}_S (\Omega^{SS}_\mathscr{A})^{-1} = \Omega^{SS}_{2\mu_1+\mu_2} (\Omega^{SS}_{2\mu_2})^{-1}$. I have verified this fact from the explicit data, but it is possible something deeper ensures its truth. The same sort of tensor-product graph and identity applies for $so(n)_k$ with $\rho=\mathscr{A}$ and $\phi=S$ as well.

\section{Conclusions}
\label{sec:conclusion}

A glib but not meaningless way of summarising integrable lattice models is as ``integrability requires adding geometry to topology''. Boltzmann weights of trigonometric integrable models involve both topological invariants and local weights. The local information depends on the spectral parameter, the angle between two lines of the lattice in \eqref{YBE} for most solutions of the Yang-Baxter equation \cite{Baxter1978}. Thinking of the conserved-current relation \eqref{Jcons} as a lattice analog of a divergence-free condition gives a natural explanation for why angles and hence geometry appear \cite{Cardy2009}. 

Requiring a conserved current exist is much easier way of finding trigonometric Boltzmann weights than solving the Yang-Baxter equation \cite{Cardy2009}. Moreover, the categorical approach described in this paper provides a simple method both to define the currents and then find the weights that make them conserved. This simplicity of \eqref{central} points the way to being able to classify which objects in a category will be able to be Baxterised. In particular, since all the data is known for categories built on quantum-group algebras, it very well may be possible to classify all such integrable models. 

Just as all known Boltzmann weights admitting a non-trivial conserved current also go on to also solve the Yang-Baxter equation, the converse also may be true. Namely, of all the known unitary trigonometric solutions to Yang-Baxter, I know of none that can {\em not} be written in terms of category data with a conserved current. Of course, knowing no counterexamples is hardly the same as knowing the truth, but it is a promising start. Given the simplicity of the construction, with a little patience it should be possible to check many more examples, and very possibly even prove (or disprove) that all unitary trigonometric solutions of the Yang-Baxter equation are of the form \eqref{central}. 

I made a number of assumptions to simplify the analysis, but none of them seem particularly crucial.  Models where $\rho$ is not simple very possibly can be obtained by reduction from simple objects. For example, the ``dilute $O(n)$'' model (where $\rho = 0 \oplus \tfrac12$ in $su(2)_k$ or $\mathcal{A}_{k+1}$ language) is related to the $A_2^{(2)}$ Izergin-Korepin solution described above \cite{Zhou1997}. Although for the most part I avoided categories with non-self-dual objects, the fact that the clock-model example worked beautifully is a good omen for extending the results to such models. Placing different objects on the horizontal and vertical strands as in \cite{Delius1994,Delius1995} and including objects where $N_{ab}^c>1$ also seem to present no major obstacle. 

Even more intriguingly, the results may be applicable directly to fusion categories. In \cite{Aasen2020}, topological defects were constructed in lattice models built on a fusion category seemingly without recourse to braiding.  Such defects come from the Drinfeld centre, a braided tensor category associated with any fusion category, even those without braiding \cite{Muger2003}. The construction 
can be extended to allow these defect lines to be terminated without braiding, and most importantly, with the appropriate behaviour under twists \cite{Aasen2020b}. 
It seems very possible that the construction in this paper can be extended to cover such cases, for example the Haagerup category and others discussed e.g.\ in \cite{Hong2008}. Even more exciting would be if such conserved currents were then to lead to integrable lattice models.

Another very interesting direction to pursue is to understand if the assumptions made here can be relaxed even further to cover various more complicated integrable models such as those built on graded Lie ``superalgebras''. Typically the associated categories are not unitary and can have an infinite number of simple objects. Nevertheless the quantum-group approach does work \cite{Gould2002}, boding well for extending the categorical approach as well. 

One glaring hole remains though. The analysis so far does not provide a way to address elliptic solutions of Yang-Baxter, where the associated lattice models are integrable but not critical. Most and possibly all trigonometric solutions admit at least one elliptic deformation, but the connection to the category is not so obvious. However, recent progress has been made by extending Chern-Simons topological field theory from three spacetime dimensions to four. Elliptic solutions to Yang-Baxter arise with the extra dimension playing the role of the spectral parameter \cite{Costello2018}. It would be quite exciting to relate this field-theory approach to the categorical one, even in the trigonometric case. 

\medskip
\paragraph{Acknowledgments} I am very grateful to David Aasen and Roger Mong for collaboration on \cite{Aasen2016,Aasen2020} and for their mentoring in the Way of the Category. I thank Denis Bernard and John Cardy for essential conversations many moons ago, and Niall Mackay, Eric Rowell and Eric Vernier for helpful comments and  guidance to the literature. This work was supported by EPSRC grants EP/S020527/1 and EP/N01930X. 



\bigskip

\setlength{\bibsep}{4.5pt plus 0.3ex}
\bibliography{currentsrefs}

\begin{thebibliography}{81}%
\makeatletter
\providecommand \@ifxundefined [1]{%
 \@ifx{#1\undefined}
}%
\providecommand \@ifnum [1]{%
 \ifnum #1\expandafter \@firstoftwo
 \else \expandafter \@secondoftwo
 \fi
}%
\providecommand \@ifx [1]{%
 \ifx #1\expandafter \@firstoftwo
 \else \expandafter \@secondoftwo
 \fi
}%
\providecommand \natexlab [1]{#1}%
\providecommand \enquote  [1]{``#1''}%
\providecommand \bibnamefont  [1]{#1}%
\providecommand \bibfnamefont [1]{#1}%
\providecommand \citenamefont [1]{#1}%
\providecommand \href@noop [0]{\@secondoftwo}%
\providecommand \href [0]{\begingroup \@sanitize@url \@href}%
\providecommand \@href[1]{\@@startlink{#1}\@@href}%
\providecommand \@@href[1]{\endgroup#1\@@endlink}%
\providecommand \@sanitize@url [0]{\catcode `\\12\catcode `\$12\catcode
  `\&12\catcode `\#12\catcode `\^12\catcode `\_12\catcode `\%12\relax}%
\providecommand \@@startlink[1]{}%
\providecommand \@@endlink[0]{}%
\providecommand \url  [0]{\begingroup\@sanitize@url \@url }%
\providecommand \@url [1]{\endgroup\@href {#1}{\urlprefix }}%
\providecommand \urlprefix  [0]{URL }%
\providecommand \Eprint [0]{\href }%
\providecommand \doibase [0]{http://dx.doi.org/}%
\providecommand \selectlanguage [0]{\@gobble}%
\providecommand \bibinfo  [0]{\@secondoftwo}%
\providecommand \bibfield  [0]{\@secondoftwo}%
\providecommand \translation [1]{[#1]}%
\providecommand \BibitemOpen [0]{}%
\providecommand \bibitemStop [0]{}%
\providecommand \bibitemNoStop [0]{.\EOS\space}%
\providecommand \EOS [0]{\spacefactor3000\relax}%
\providecommand \BibitemShut  [1]{\csname bibitem#1\endcsname}%
\let\auto@bib@innerbib\@empty
\bibitem [{\citenamefont {Temperley}\ and\ \citenamefont
  {Lieb}(1971)}]{Temperley1971}%
  \BibitemOpen
  \bibfield  {author} {\bibinfo {author} {\bibfnamefont {H.~N.~V.}\
  \bibnamefont {Temperley}}\ and\ \bibinfo {author} {\bibfnamefont {E.~H.}\
  \bibnamefont {Lieb}},\ }\href {\doibase 10.1098/rspa.1971.0067} {\bibfield
  {journal} {\bibinfo  {journal} {Proc. Roy. Soc. Lond.}\ }\textbf {\bibinfo
  {volume} {A322}},\ \bibinfo {pages} {251} (\bibinfo {year}
  {1971})}\BibitemShut {NoStop}%
\bibitem [{\citenamefont {Fortuin}\ and\ \citenamefont
  {Kasteleyn}(1972)}]{Fortuin1971}%
  \BibitemOpen
  \bibfield  {author} {\bibinfo {author} {\bibfnamefont {C.~M.}\ \bibnamefont
  {Fortuin}}\ and\ \bibinfo {author} {\bibfnamefont {P.~W.}\ \bibnamefont
  {Kasteleyn}},\ }\href {\doibase 10.1016/0031-8914(72)90045-6} {\bibfield
  {journal} {\bibinfo  {journal} {Physica}\ }\textbf {\bibinfo {volume} {57}},\
  \bibinfo {pages} {536} (\bibinfo {year} {1972})}\BibitemShut {NoStop}%
\bibitem [{\citenamefont {Baxter}\ \emph {et~al.}(1976)\citenamefont {Baxter},
  \citenamefont {Kelland},\ and\ \citenamefont {Wu}}]{Baxter1976}%
  \BibitemOpen
  \bibfield  {author} {\bibinfo {author} {\bibfnamefont {R.~J.}\ \bibnamefont
  {Baxter}}, \bibinfo {author} {\bibfnamefont {S.~B.}\ \bibnamefont {Kelland}},
  \ and\ \bibinfo {author} {\bibfnamefont {F.~Y.}\ \bibnamefont {Wu}},\ }\href
  {\doibase 10.1088/0305-4470/9/3/009} {\bibfield  {journal} {\bibinfo
  {journal} {J. Phys. A}\ }\textbf {\bibinfo {volume} {9}},\ \bibinfo {pages}
  {397} (\bibinfo {year} {1976})}\BibitemShut {NoStop}%
\bibitem [{\citenamefont {Andrews}\ \emph {et~al.}(1984)\citenamefont
  {Andrews}, \citenamefont {Baxter},\ and\ \citenamefont
  {Forrester}}]{Andrews1984}%
  \BibitemOpen
  \bibfield  {author} {\bibinfo {author} {\bibfnamefont {G.}~\bibnamefont
  {Andrews}}, \bibinfo {author} {\bibfnamefont {R.}~\bibnamefont {Baxter}}, \
  and\ \bibinfo {author} {\bibfnamefont {P.}~\bibnamefont {Forrester}},\ }\href
  {\doibase 10.1007/BF01014383} {\bibfield  {journal} {\bibinfo  {journal} {J.\
  Stat.\ Phys.}\ }\textbf {\bibinfo {volume} {35}},\ \bibinfo {pages} {193}
  (\bibinfo {year} {1984})}\BibitemShut {NoStop}%
\bibitem [{\citenamefont {Pasquier}(1987)}]{Pasquier1986}%
  \BibitemOpen
  \bibfield  {author} {\bibinfo {author} {\bibfnamefont {V.}~\bibnamefont
  {Pasquier}},\ }\href {\doibase 10.1016/0550-3213(87)90332-4} {\bibfield
  {journal} {\bibinfo  {journal} {Nucl. Phys.}\ }\textbf {\bibinfo {volume}
  {B285}},\ \bibinfo {pages} {162} (\bibinfo {year} {1987})}\BibitemShut
  {NoStop}%
\bibitem [{\citenamefont {Birman}\ and\ \citenamefont
  {Wenzl}(1989)}]{Birman1989}%
  \BibitemOpen
  \bibfield  {author} {\bibinfo {author} {\bibfnamefont {J.~S.}\ \bibnamefont
  {Birman}}\ and\ \bibinfo {author} {\bibfnamefont {H.}~\bibnamefont {Wenzl}},\
  }\href@noop {} {\bibfield  {journal} {\bibinfo  {journal} {Trans. Am. Math.
  Soc.}\ }\textbf {\bibinfo {volume} {313}},\ \bibinfo {pages} {249} (\bibinfo
  {year} {1989})}\BibitemShut {NoStop}%
\bibitem [{\citenamefont {Murakami}(1990)}]{Murakami1990}%
  \BibitemOpen
  \bibfield  {author} {\bibinfo {author} {\bibfnamefont {J.}~\bibnamefont
  {Murakami}},\ }in\ \href@noop {} {\emph {\bibinfo {booktitle} {New
  Developments In The Theory Of Knots}}}\ (\bibinfo  {publisher} {World
  Scientific},\ \bibinfo {year} {1990})\ pp.\ \bibinfo {pages}
  {480--493}\BibitemShut {NoStop}%
\bibitem [{\citenamefont {Jimbo}\ \emph {et~al.}(1988)\citenamefont {Jimbo},
  \citenamefont {Miwa},\ and\ \citenamefont {Okado}}]{Jimbo1988}%
  \BibitemOpen
  \bibfield  {author} {\bibinfo {author} {\bibfnamefont {M.}~\bibnamefont
  {Jimbo}}, \bibinfo {author} {\bibfnamefont {T.}~\bibnamefont {Miwa}}, \ and\
  \bibinfo {author} {\bibfnamefont {M.}~\bibnamefont {Okado}},\ }\href
  {\doibase 10.1007/BF01229206} {\bibfield  {journal} {\bibinfo  {journal}
  {Commun. Math. Phys.}\ }\textbf {\bibinfo {volume} {116}},\ \bibinfo {pages}
  {507} (\bibinfo {year} {1988})}\BibitemShut {NoStop}%
\bibitem [{\citenamefont {Jones}(1987)}]{Jones1987}%
  \BibitemOpen
  \bibfield  {author} {\bibinfo {author} {\bibfnamefont {V.}~\bibnamefont
  {Jones}},\ }\href {\doibase 10.2307/1971403} {\bibfield  {journal} {\bibinfo
  {journal} {Annals Math.}\ }\textbf {\bibinfo {volume} {126}},\ \bibinfo
  {pages} {335} (\bibinfo {year} {1987})}\BibitemShut {NoStop}%
\bibitem [{\citenamefont {Kauffman}(1991)}]{Kauffman1991}%
  \BibitemOpen
  \bibfield  {author} {\bibinfo {author} {\bibfnamefont {L.}~\bibnamefont
  {Kauffman}},\ }\href {https://books.google.co.uk/books?id=av05vRwIKIwC}
  {\emph {\bibinfo {title} {Knots and Physics}}},\ K \& E series on knots and
  everything\ (\bibinfo  {publisher} {World Scientific},\ \bibinfo {year}
  {1991})\BibitemShut {NoStop}%
\bibitem [{\citenamefont {Moore}\ and\ \citenamefont
  {Seiberg}(1989)}]{Moore1989}%
  \BibitemOpen
  \bibfield  {author} {\bibinfo {author} {\bibfnamefont {G.~W.}\ \bibnamefont
  {Moore}}\ and\ \bibinfo {author} {\bibfnamefont {N.}~\bibnamefont
  {Seiberg}},\ }in\ \href {http://alice.cern.ch/format/showfull?sysnb=0113749}
  {\emph {\bibinfo {booktitle} {{1989 Banff NATO ASI: Physics, Geometry and
  Topology Banff, Canada, August 14-25, 1989}}}}\ (\bibinfo {year}
  {1989})\BibitemShut {NoStop}%
\bibitem [{\citenamefont {Kitaev}(2006)}]{Kitaev2006}%
  \BibitemOpen
  \bibfield  {author} {\bibinfo {author} {\bibfnamefont {A.}~\bibnamefont
  {Kitaev}},\ }\href {\doibase 10.1016/j.aop.2005.10.005} {\bibfield  {journal}
  {\bibinfo  {journal} {Annals Phys.}\ }\textbf {\bibinfo {volume} {321}},\
  \bibinfo {pages} {2} (\bibinfo {year} {2006})}\BibitemShut {NoStop}%
\bibitem [{\citenamefont {Bonderson}(2007)}]{Bondersonthesis}%
  \BibitemOpen
  \bibfield  {author} {\bibinfo {author} {\bibfnamefont {P.~H.}\ \bibnamefont
  {Bonderson}},\ }\emph {\bibinfo {title} {{Non-abelian anyons and
  interferometry}}},\ \href
  {http://etd.caltech.edu/etd/available/etd-06042007-101617/unrestricted/thesis.pdf}
  {Ph.D. thesis},\ \bibinfo  {school} {Caltech} (\bibinfo {year}
  {2007})\BibitemShut {NoStop}%
\bibitem [{\citenamefont {{Aasen}}\ \emph {et~al.}(2020)\citenamefont
  {{Aasen}}, \citenamefont {Fendley},\ and\ \citenamefont {Mong}}]{Aasen2020}%
  \BibitemOpen
  \bibfield  {author} {\bibinfo {author} {\bibfnamefont {D.}~\bibnamefont
  {{Aasen}}}, \bibinfo {author} {\bibfnamefont {P.}~\bibnamefont {Fendley}}, \
  and\ \bibinfo {author} {\bibfnamefont {R.}~\bibnamefont {Mong}},\ }\href@noop
  {} {\enquote {\bibinfo {title} {Topological Defects on the Lattice: Dualities
  and Degeneracies},}\ } (\bibinfo {year} {2020}),\ \bibinfo {note} {to appear
  soon}\BibitemShut {NoStop}%
\bibitem [{\citenamefont {Baxter}(1982)}]{Baxter1982}%
  \BibitemOpen
  \bibfield  {author} {\bibinfo {author} {\bibfnamefont {R.~J.}\ \bibnamefont
  {Baxter}},\ }\href@noop {} {\emph {\bibinfo {title} {{Exactly solved models
  in statistical mechanics}}}}\ (\bibinfo  {publisher} {Academic},\ \bibinfo
  {year} {1982})\BibitemShut {NoStop}%
\bibitem [{\citenamefont {Jones}(1989)}]{Jones1989}%
  \BibitemOpen
  \bibfield  {author} {\bibinfo {author} {\bibfnamefont {V.~F.~R.}\
  \bibnamefont {Jones}},\ }\href
  {https://projecteuclid.org:443/euclid.pjm/1102650387} {\bibfield  {journal}
  {\bibinfo  {journal} {Pacific J. Math.}\ }\textbf {\bibinfo {volume} {137}},\
  \bibinfo {pages} {311} (\bibinfo {year} {1989})}\BibitemShut {NoStop}%
\bibitem [{\citenamefont {Wadati}\ \emph {et~al.}(1989)\citenamefont {Wadati},
  \citenamefont {Deguchi},\ and\ \citenamefont {Akutsu}}]{Wadati1989}%
  \BibitemOpen
  \bibfield  {author} {\bibinfo {author} {\bibfnamefont {M.}~\bibnamefont
  {Wadati}}, \bibinfo {author} {\bibfnamefont {T.}~\bibnamefont {Deguchi}}, \
  and\ \bibinfo {author} {\bibfnamefont {Y.}~\bibnamefont {Akutsu}},\ }\href
  {\doibase 10.1016/0370-1573(89)90123-3} {\bibfield  {journal} {\bibinfo
  {journal} {Phys. Rept.}\ }\textbf {\bibinfo {volume} {180}},\ \bibinfo
  {pages} {247} (\bibinfo {year} {1989})}\BibitemShut {NoStop}%
\bibitem [{\citenamefont {Wu}(1992)}]{Wu1992}%
  \BibitemOpen
  \bibfield  {author} {\bibinfo {author} {\bibfnamefont {F.~Y.}\ \bibnamefont
  {Wu}},\ }\href {\doibase 10.1103/RevModPhys.64.1099} {\bibfield  {journal}
  {\bibinfo  {journal} {Rev. Mod. Phys.}\ }\textbf {\bibinfo {volume} {64}},\
  \bibinfo {pages} {1099} (\bibinfo {year} {1992})},\ \bibinfo {note}
  {[Erratum: Rev. Mod. Phys.65,577(1993)]}\BibitemShut {NoStop}%
\bibitem [{\citenamefont {Jones}(1990)}]{Jones1990}%
  \BibitemOpen
  \bibfield  {author} {\bibinfo {author} {\bibfnamefont {V.~F.~R.}\
  \bibnamefont {Jones}},\ }\enquote {\bibinfo {title} {Baxterization},}\ in\
  \href {\doibase 10.1007/978-1-4684-9148-7_2} {\emph {\bibinfo {booktitle}
  {Differential Geometric Methods in Theoretical Physics: Physics and
  Geometry}}}\ (\bibinfo  {publisher} {Springer US},\ \bibinfo {address}
  {Boston, MA},\ \bibinfo {year} {1990})\ pp.\ \bibinfo {pages}
  {5--11}\BibitemShut {NoStop}%
\bibitem [{\citenamefont {Jones}(2003)}]{Jones2003}%
  \BibitemOpen
  \bibfield  {author} {\bibinfo {author} {\bibfnamefont {V.~F.}\ \bibnamefont
  {Jones}},\ }\href@noop {} {\enquote {\bibinfo {title} {In and around the
  origin of quantum groups},}\ } (\bibinfo {year} {2003}),\ \Eprint
  {http://arxiv.org/abs/1805.05736} {arXiv:1805.05736} \BibitemShut {NoStop}%
\bibitem [{\citenamefont {Drinfeld}(1985)}]{Drinfeld1985}%
  \BibitemOpen
  \bibfield  {author} {\bibinfo {author} {\bibfnamefont {V.}~\bibnamefont
  {Drinfeld}},\ }\href@noop {} {\bibfield  {journal} {\bibinfo  {journal} {Sov.
  Math. Dokl.}\ }\textbf {\bibinfo {volume} {32}},\ \bibinfo {pages} {254}
  (\bibinfo {year} {1985})}\BibitemShut {NoStop}%
\bibitem [{\citenamefont {Kirillov}\ and\ \citenamefont
  {Reshetikhin}(1990)}]{Kirillov1990}%
  \BibitemOpen
  \bibfield  {author} {\bibinfo {author} {\bibfnamefont {A.}~\bibnamefont
  {Kirillov}}\ and\ \bibinfo {author} {\bibfnamefont {N.}~\bibnamefont
  {Reshetikhin}},\ }\href {\doibase 10.1007/BF02097710} {\bibfield  {journal}
  {\bibinfo  {journal} {Commun. Math. Phys.}\ }\textbf {\bibinfo {volume}
  {134}},\ \bibinfo {pages} {421} (\bibinfo {year} {1990})}\BibitemShut
  {NoStop}%
\bibitem [{\citenamefont {{Khoroshkin}}\ and\ \citenamefont
  {{Tolstoy}}(1991)}]{Khoroshkin1991}%
  \BibitemOpen
  \bibfield  {author} {\bibinfo {author} {\bibfnamefont {S.~M.}\ \bibnamefont
  {{Khoroshkin}}}\ and\ \bibinfo {author} {\bibfnamefont {V.~N.}\ \bibnamefont
  {{Tolstoy}}},\ }\href {\doibase 10.1007/BF02102819} {\bibfield  {journal}
  {\bibinfo  {journal} {Commun. Math. Phys.}\ }\textbf {\bibinfo {volume}
  {141}},\ \bibinfo {pages} {599} (\bibinfo {year} {1991})}\BibitemShut
  {NoStop}%
\bibitem [{\citenamefont {Gomez}\ \emph {et~al.}(2011)\citenamefont {Gomez},
  \citenamefont {Sierra},\ and\ \citenamefont {Ruiz-Altaba}}]{Gomez1996}%
  \BibitemOpen
  \bibfield  {author} {\bibinfo {author} {\bibfnamefont {C.}~\bibnamefont
  {Gomez}}, \bibinfo {author} {\bibfnamefont {G.}~\bibnamefont {Sierra}}, \
  and\ \bibinfo {author} {\bibfnamefont {M.}~\bibnamefont {Ruiz-Altaba}},\
  }\href {\doibase 10.1017/CBO9780511628825} {\emph {\bibinfo {title} {{Quantum
  groups in two-dimensional physics}}}},\ Cambridge Monographs on Mathematical
  Physics\ (\bibinfo  {publisher} {Cambridge University Press},\ \bibinfo
  {year} {2011})\BibitemShut {NoStop}%
\bibitem [{\citenamefont {Jimbo}(1986)}]{Jimbo1986}%
  \BibitemOpen
  \bibfield  {author} {\bibinfo {author} {\bibfnamefont {M.}~\bibnamefont
  {Jimbo}},\ }\href {\doibase 10.1007/BF01221646} {\bibfield  {journal}
  {\bibinfo  {journal} {Commun. Math. Phys.}\ }\textbf {\bibinfo {volume}
  {102}},\ \bibinfo {pages} {537} (\bibinfo {year} {1986})}\BibitemShut
  {NoStop}%
\bibitem [{\citenamefont {Zhang}\ \emph {et~al.}(1991)\citenamefont {Zhang},
  \citenamefont {Gould},\ and\ \citenamefont {Bracken}}]{Zhang1990}%
  \BibitemOpen
  \bibfield  {author} {\bibinfo {author} {\bibfnamefont {R.-b.}\ \bibnamefont
  {Zhang}}, \bibinfo {author} {\bibfnamefont {M.}~\bibnamefont {Gould}}, \ and\
  \bibinfo {author} {\bibfnamefont {A.}~\bibnamefont {Bracken}},\ }\href
  {\doibase 10.1016/0550-3213(91)90369-9} {\bibfield  {journal} {\bibinfo
  {journal} {Nucl. Phys. B}\ }\textbf {\bibinfo {volume} {354}},\ \bibinfo
  {pages} {625} (\bibinfo {year} {1991})}\BibitemShut {NoStop}%
\bibitem [{\citenamefont {Delius}\ \emph {et~al.}(1994)\citenamefont {Delius},
  \citenamefont {Gould},\ and\ \citenamefont {Zhang}}]{Delius1994}%
  \BibitemOpen
  \bibfield  {author} {\bibinfo {author} {\bibfnamefont {G.~W.}\ \bibnamefont
  {Delius}}, \bibinfo {author} {\bibfnamefont {M.~D.}\ \bibnamefont {Gould}}, \
  and\ \bibinfo {author} {\bibfnamefont {Y.-Z.}\ \bibnamefont {Zhang}},\ }\href
  {\doibase 10.1016/0550-3213(94)90607-6} {\bibfield  {journal} {\bibinfo
  {journal} {Nucl. Phys. B}\ }\textbf {\bibinfo {volume} {432}},\ \bibinfo
  {pages} {377} (\bibinfo {year} {1994})},\ \Eprint
  {http://arxiv.org/abs/hep-th/9405030} {arXiv:hep-th/9405030} \BibitemShut
  {NoStop}%
\bibitem [{\citenamefont {{Delius}}\ \emph {et~al.}(1996)\citenamefont
  {{Delius}}, \citenamefont {{Gould}},\ and\ \citenamefont
  {{Zhang}}}]{Delius1995}%
  \BibitemOpen
  \bibfield  {author} {\bibinfo {author} {\bibfnamefont {G.~W.}\ \bibnamefont
  {{Delius}}}, \bibinfo {author} {\bibfnamefont {M.~D.}\ \bibnamefont
  {{Gould}}}, \ and\ \bibinfo {author} {\bibfnamefont {Y.-Z.}\ \bibnamefont
  {{Zhang}}},\ }\href {\doibase 10.1142/S0217751X96001632} {\bibfield
  {journal} {\bibinfo  {journal} {Int. J. Mod. Phys. A}\ }\textbf {\bibinfo
  {volume} {11}},\ \bibinfo {pages} {3415} (\bibinfo {year} {1996})},\ \Eprint
  {http://arxiv.org/abs/q-alg/9508012} {arXiv:q-alg/9508012} \BibitemShut
  {NoStop}%
\bibitem [{\citenamefont {Bernard}\ and\ \citenamefont
  {Felder}(1991)}]{Bernard1991}%
  \BibitemOpen
  \bibfield  {author} {\bibinfo {author} {\bibfnamefont {D.}~\bibnamefont
  {Bernard}}\ and\ \bibinfo {author} {\bibfnamefont {G.}~\bibnamefont
  {Felder}},\ }\href {\doibase 10.1016/0550-3213(91)90608-Z} {\bibfield
  {journal} {\bibinfo  {journal} {Nucl. Phys.}\ }\textbf {\bibinfo {volume}
  {B365}},\ \bibinfo {pages} {98} (\bibinfo {year} {1991})}\BibitemShut
  {NoStop}%
\bibitem [{\citenamefont {Smirnov}(2006)}]{Smirnov2006}%
  \BibitemOpen
  \bibfield  {author} {\bibinfo {author} {\bibfnamefont {S.}~\bibnamefont
  {Smirnov}},\ }\href@noop {} {\bibfield  {journal} {\bibinfo  {journal}
  {Proc.Int.Congr.Math.}\ }\textbf {\bibinfo {volume} {2}},\ \bibinfo {pages}
  {1421} (\bibinfo {year} {2006})},\ \Eprint {http://arxiv.org/abs/0708.0032}
  {arXiv:0708.0032} \BibitemShut {NoStop}%
\bibitem [{\citenamefont {{Cardy}}(2009)}]{Cardy2009}%
  \BibitemOpen
  \bibfield  {author} {\bibinfo {author} {\bibfnamefont {J.}~\bibnamefont
  {{Cardy}}},\ }\href {\doibase 10.1007/s10955-009-9870-6} {\bibfield
  {journal} {\bibinfo  {journal} {J. Stat. Phys.}\ }\textbf {\bibinfo {volume}
  {137}},\ \bibinfo {pages} {814} (\bibinfo {year} {2009})},\ \Eprint
  {http://arxiv.org/abs/0907.4070} {arXiv:0907.4070} \BibitemShut {NoStop}%
\bibitem [{\citenamefont {Shankar}\ and\ \citenamefont
  {Witten}(1978)}]{Shankar77}%
  \BibitemOpen
  \bibfield  {author} {\bibinfo {author} {\bibfnamefont {R.}~\bibnamefont
  {Shankar}}\ and\ \bibinfo {author} {\bibfnamefont {E.}~\bibnamefont
  {Witten}},\ }\href {\doibase 10.1103/PhysRevD.17.2134} {\bibfield  {journal}
  {\bibinfo  {journal} {Phys. Rev.}\ }\textbf {\bibinfo {volume} {D17}},\
  \bibinfo {pages} {2134} (\bibinfo {year} {1978})}\BibitemShut {NoStop}%
\bibitem [{\citenamefont {Dorey}(1996)}]{Dorey97}%
  \BibitemOpen
  \bibfield  {author} {\bibinfo {author} {\bibfnamefont {P.}~\bibnamefont
  {Dorey}},\ }in\ \href@noop {} {\emph {\bibinfo {booktitle} {{Eotvos Summer
  School in Physics: Conformal Field Theories and Integrable Models}}}}\
  (\bibinfo {year} {1996})\ pp.\ \bibinfo {pages} {85--125},\ \Eprint
  {http://arxiv.org/abs/hep-th/9810026} {arXiv:hep-th/9810026} \BibitemShut
  {NoStop}%
\bibitem [{\citenamefont {Bakalov}\ and\ \citenamefont
  {Kirillov}(2001)}]{Bakalov2001}%
  \BibitemOpen
  \bibfield  {author} {\bibinfo {author} {\bibfnamefont {B.}~\bibnamefont
  {Bakalov}}\ and\ \bibinfo {author} {\bibfnamefont {A.}~\bibnamefont
  {Kirillov}},\ }\href {\doibase http://dx.doi.org/10.1090/ulect/021} {\emph
  {\bibinfo {title} {Lectures on tensor categories and modular functors}}},\
  Vol.~\bibinfo {volume} {21}\ (\bibinfo {year} {2001})\BibitemShut {NoStop}%
\bibitem [{\citenamefont {Walker}(2006)}]{Walker2006}%
  \BibitemOpen
  \bibfield  {author} {\bibinfo {author} {\bibfnamefont {K.}~\bibnamefont
  {Walker}},\ }\href {http://canyon23.net/math/} {\emph {\bibinfo {title}
  {TQFTs}}}\ (\bibinfo {year} {2006})\BibitemShut {NoStop}%
\bibitem [{\citenamefont {Reshetikhin}(1988)}]{Reshetikhin1988}%
  \BibitemOpen
  \bibfield  {author} {\bibinfo {author} {\bibfnamefont {N.}~\bibnamefont
  {Reshetikhin}},\ }\href@noop {} {\  (\bibinfo {year} {1988})},\ \bibinfo
  {note} {{unpublished manuscripts, {\it Quantized universal enveloping
  algebras, the Yang-Baxter equation and invariants of links}, I and
  II}}\BibitemShut {NoStop}%
\bibitem [{\citenamefont {Turaev}(1992)}]{Turaev1992}%
  \BibitemOpen
  \bibfield  {author} {\bibinfo {author} {\bibfnamefont {V.~G.}\ \bibnamefont
  {Turaev}},\ }\href {http://projecteuclid.org/euclid.jdg/1214448442}
  {\bibfield  {journal} {\bibinfo  {journal} {J. Differential Geom.}\ }\textbf
  {\bibinfo {volume} {36}},\ \bibinfo {pages} {35} (\bibinfo {year}
  {1992})}\BibitemShut {NoStop}%
\bibitem [{\citenamefont {Turaev}\ and\ \citenamefont
  {Viro}(1992)}]{TuraevViro}%
  \BibitemOpen
  \bibfield  {author} {\bibinfo {author} {\bibfnamefont {V.}~\bibnamefont
  {Turaev}}\ and\ \bibinfo {author} {\bibfnamefont {O.}~\bibnamefont {Viro}},\
  }\href {\doibase 10.1016/0040-9383(92)90015-A} {\bibfield  {journal}
  {\bibinfo  {journal} {Topology}\ }\textbf {\bibinfo {volume} {31}},\ \bibinfo
  {pages} {865} (\bibinfo {year} {1992})}\BibitemShut {NoStop}%
\bibitem [{\citenamefont {Barrett}\ and\ \citenamefont
  {Westbury}(1996)}]{Barrett1996}%
  \BibitemOpen
  \bibfield  {author} {\bibinfo {author} {\bibfnamefont {J.}~\bibnamefont
  {Barrett}}\ and\ \bibinfo {author} {\bibfnamefont {B.}~\bibnamefont
  {Westbury}},\ }\href {\doibase 10.1090/S0002-9947-96-01660-1} {\bibfield
  {journal} {\bibinfo  {journal} {Trans. Am. Math. Soc.}\ }\textbf {\bibinfo
  {volume} {348}},\ \bibinfo {pages} {3997} (\bibinfo {year} {1996})},\ \Eprint
  {http://arxiv.org/abs/hep-th/9311155} {arXiv:hep-th/9311155} \BibitemShut
  {NoStop}%
\bibitem [{\citenamefont {Feiguin}\ \emph {et~al.}(2007)\citenamefont
  {Feiguin}, \citenamefont {Trebst}, \citenamefont {Ludwig}, \citenamefont
  {Troyer}, \citenamefont {Kitaev}, \citenamefont {Wang},\ and\ \citenamefont
  {Freedman}}]{Feiguin2007}%
  \BibitemOpen
  \bibfield  {author} {\bibinfo {author} {\bibfnamefont {A.}~\bibnamefont
  {Feiguin}}, \bibinfo {author} {\bibfnamefont {S.}~\bibnamefont {Trebst}},
  \bibinfo {author} {\bibfnamefont {A.~W.~W.}\ \bibnamefont {Ludwig}}, \bibinfo
  {author} {\bibfnamefont {M.}~\bibnamefont {Troyer}}, \bibinfo {author}
  {\bibfnamefont {A.}~\bibnamefont {Kitaev}}, \bibinfo {author} {\bibfnamefont
  {Z.}~\bibnamefont {Wang}}, \ and\ \bibinfo {author} {\bibfnamefont {M.~H.}\
  \bibnamefont {Freedman}},\ }\href {\doibase 10.1103/PhysRevLett.98.160409}
  {\bibfield  {journal} {\bibinfo  {journal} {Phys. Rev. Lett.}\ }\textbf
  {\bibinfo {volume} {98}},\ \bibinfo {pages} {160409} (\bibinfo {year}
  {2007})},\ \Eprint {http://arxiv.org/abs/cond-mat/0612341}
  {arXiv:cond-mat/0612341} \BibitemShut {NoStop}%
\bibitem [{\citenamefont {Au-Yang}\ \emph {et~al.}(1987)\citenamefont
  {Au-Yang}, \citenamefont {McCoy}, \citenamefont {Perk}, \citenamefont
  {Tang},\ and\ \citenamefont {Yan}}]{AuYang1987}%
  \BibitemOpen
  \bibfield  {author} {\bibinfo {author} {\bibfnamefont {H.}~\bibnamefont
  {Au-Yang}}, \bibinfo {author} {\bibfnamefont {B.~M.}\ \bibnamefont {McCoy}},
  \bibinfo {author} {\bibfnamefont {J.~H.~H.}\ \bibnamefont {Perk}}, \bibinfo
  {author} {\bibfnamefont {S.}~\bibnamefont {Tang}}, \ and\ \bibinfo {author}
  {\bibfnamefont {M.-L.}\ \bibnamefont {Yan}},\ }\href {\doibase
  http://dx.doi.org/10.1016/0375-9601(87)90065-X} {\bibfield  {journal}
  {\bibinfo  {journal} {Phys. Lett. A}\ }\textbf {\bibinfo {volume} {123}},\
  \bibinfo {pages} {219 } (\bibinfo {year} {1987})}\BibitemShut {NoStop}%
\bibitem [{\citenamefont {Aasen}\ \emph {et~al.}(2016)\citenamefont {Aasen},
  \citenamefont {Mong},\ and\ \citenamefont {Fendley}}]{Aasen2016}%
  \BibitemOpen
  \bibfield  {author} {\bibinfo {author} {\bibfnamefont {D.}~\bibnamefont
  {Aasen}}, \bibinfo {author} {\bibfnamefont {R.~S.~K.}\ \bibnamefont {Mong}},
  \ and\ \bibinfo {author} {\bibfnamefont {P.}~\bibnamefont {Fendley}},\ }\href
  {\doibase 10.1088/1751-8113/49/35/354001} {\bibfield  {journal} {\bibinfo
  {journal} {J. Phys. A}\ }\textbf {\bibinfo {volume} {49}},\ \bibinfo {pages}
  {354001} (\bibinfo {year} {2016})},\ \Eprint
  {http://arxiv.org/abs/1601.07185} {arXiv:1601.07185} \BibitemShut {NoStop}%
\bibitem [{\citenamefont {Birman}\ and\ \citenamefont
  {Cannon}(1974)}]{Birman1975}%
  \BibitemOpen
  \bibfield  {author} {\bibinfo {author} {\bibfnamefont {J.~S.}\ \bibnamefont
  {Birman}}\ and\ \bibinfo {author} {\bibfnamefont {J.}~\bibnamefont
  {Cannon}},\ }\href {\doibase http://dx.doi.org/10.2307/j.ctt1b9rzv3} {\emph
  {\bibinfo {title} {Braids, Links, and Mapping Class Groups. (AM-82)}}}\
  (\bibinfo  {publisher} {Princeton University Press},\ \bibinfo {year}
  {1974})\BibitemShut {NoStop}%
\bibitem [{\citenamefont {Magnus}(1974)}]{Magnus1974}%
  \BibitemOpen
  \bibfield  {author} {\bibinfo {author} {\bibfnamefont {W.}~\bibnamefont
  {Magnus}},\ }in\ \href {\doibase 10.1007/978-3-662-21571-5_49} {\emph
  {\bibinfo {booktitle} {Proceedings of the Second International Conference on
  the Theory of Groups}}},\ \bibinfo {editor} {edited by\ \bibinfo {editor}
  {\bibfnamefont {M.~F.}\ \bibnamefont {Newman}}}\ (\bibinfo  {publisher}
  {Springer Berlin Heidelberg},\ \bibinfo {address} {Berlin, Heidelberg},\
  \bibinfo {year} {1974})\ pp.\ \bibinfo {pages} {463--487}\BibitemShut
  {NoStop}%
\bibitem [{\citenamefont {Belavin}\ \emph {et~al.}(1984)\citenamefont
  {Belavin}, \citenamefont {Polyakov},\ and\ \citenamefont
  {Zamolodchikov}}]{Belavin84}%
  \BibitemOpen
  \bibfield  {author} {\bibinfo {author} {\bibfnamefont {A.~A.}\ \bibnamefont
  {Belavin}}, \bibinfo {author} {\bibfnamefont {A.~M.}\ \bibnamefont
  {Polyakov}}, \ and\ \bibinfo {author} {\bibfnamefont {A.~B.}\ \bibnamefont
  {Zamolodchikov}},\ }\href@noop {} {\bibfield  {journal} {\bibinfo  {journal}
  {Nucl. Phys.}\ }\textbf {\bibinfo {volume} {B241}},\ \bibinfo {pages} {333}
  (\bibinfo {year} {1984})}\BibitemShut {NoStop}%
\bibitem [{\citenamefont {Friedan}\ \emph {et~al.}(1984)\citenamefont
  {Friedan}, \citenamefont {Qiu},\ and\ \citenamefont {Shenker}}]{Friedan1984}%
  \BibitemOpen
  \bibfield  {author} {\bibinfo {author} {\bibfnamefont {D.}~\bibnamefont
  {Friedan}}, \bibinfo {author} {\bibfnamefont {Z.}~\bibnamefont {Qiu}}, \ and\
  \bibinfo {author} {\bibfnamefont {S.}~\bibnamefont {Shenker}},\ }\href
  {\doibase 10.1103/PhysRevLett.52.1575} {\bibfield  {journal} {\bibinfo
  {journal} {Phys. Rev. Lett.}\ }\textbf {\bibinfo {volume} {52}},\ \bibinfo
  {pages} {1575} (\bibinfo {year} {1984})}\BibitemShut {NoStop}%
\bibitem [{\citenamefont {Duminil-Copin}\ and\ \citenamefont
  {Smirnov}(2012)}]{Duminil2012}%
  \BibitemOpen
  \bibfield  {author} {\bibinfo {author} {\bibfnamefont {H.}~\bibnamefont
  {Duminil-Copin}}\ and\ \bibinfo {author} {\bibfnamefont {S.}~\bibnamefont
  {Smirnov}},\ }in\ \href@noop {} {\emph {\bibinfo {booktitle} {Probability and
  Statistical Physics in Two and More Dimensions, Clay Mathematics
  Proceedings}}},\ Vol.~\bibinfo {volume} {15}\ (\bibinfo {year} {2012})\ pp.\
  \bibinfo {pages} {213--276},\ \Eprint {http://arxiv.org/abs/1109.1549}
  {arXiv:1109.1549} \BibitemShut {NoStop}%
\bibitem [{\citenamefont {Chelkak}\ and\ \citenamefont
  {Smirnov}(2012)}]{Chelkak2012}%
  \BibitemOpen
  \bibfield  {author} {\bibinfo {author} {\bibfnamefont {D.}~\bibnamefont
  {Chelkak}}\ and\ \bibinfo {author} {\bibfnamefont {S.}~\bibnamefont
  {Smirnov}},\ }\href {\doibase 10.1007/s00222-011-0371-2} {\bibfield
  {journal} {\bibinfo  {journal} {Invent. Math.}\ }\textbf {\bibinfo {volume}
  {189}},\ \bibinfo {pages} {515} (\bibinfo {year} {2012})},\ \Eprint
  {http://arxiv.org/abs/0910.2045} {arXiv:0910.2045} \BibitemShut {NoStop}%
\bibitem [{\citenamefont {Chelkak}\ \emph {et~al.}(2020)\citenamefont
  {Chelkak}, \citenamefont {Laslier},\ and\ \citenamefont
  {Russkikh}}]{Chelkak2020}%
  \BibitemOpen
  \bibfield  {author} {\bibinfo {author} {\bibfnamefont {D.}~\bibnamefont
  {Chelkak}}, \bibinfo {author} {\bibfnamefont {B.}~\bibnamefont {Laslier}}, \
  and\ \bibinfo {author} {\bibfnamefont {M.}~\bibnamefont {Russkikh}},\
  }\href@noop {} {\enquote {\bibinfo {title} {Dimer model and holomorphic
  functions on t-embeddings of planar graphs},}\ } (\bibinfo {year} {2020}),\
  \Eprint {http://arxiv.org/abs/2001.11871} {arXiv:2001.11871} \BibitemShut
  {NoStop}%
\bibitem [{\citenamefont {Fradkin}\ and\ \citenamefont
  {Kadanoff}(1980)}]{Fradkin80}%
  \BibitemOpen
  \bibfield  {author} {\bibinfo {author} {\bibfnamefont {E.}~\bibnamefont
  {Fradkin}}\ and\ \bibinfo {author} {\bibfnamefont {L.~P.}\ \bibnamefont
  {Kadanoff}},\ }\href@noop {} {\bibfield  {journal} {\bibinfo  {journal}
  {Nucl. Phys. B}\ }\textbf {\bibinfo {volume} {170}},\ \bibinfo {pages} {1}
  (\bibinfo {year} {1980})}\BibitemShut {NoStop}%
\bibitem [{\citenamefont {Riva}\ and\ \citenamefont {Cardy}(2006)}]{Riva06}%
  \BibitemOpen
  \bibfield  {author} {\bibinfo {author} {\bibfnamefont {V.}~\bibnamefont
  {Riva}}\ and\ \bibinfo {author} {\bibfnamefont {J.}~\bibnamefont {Cardy}},\
  }\href {\doibase 10.1088/1742-5468/2006/12/P12001} {\bibfield  {journal}
  {\bibinfo  {journal} {J. Stat. Mech.}\ }\textbf {\bibinfo {volume} {0612}},\
  \bibinfo {pages} {P12001} (\bibinfo {year} {2006})},\ \Eprint
  {http://arxiv.org/abs/cond-mat/0608496} {arXiv:cond-mat/0608496} \BibitemShut
  {NoStop}%
\bibitem [{\citenamefont {Mong}\ \emph {et~al.}(2014)\citenamefont {Mong},
  \citenamefont {Clarke}, \citenamefont {Alicea}, \citenamefont {Lindner},\
  and\ \citenamefont {Fendley}}]{Mong2014}%
  \BibitemOpen
  \bibfield  {author} {\bibinfo {author} {\bibfnamefont {R.~S.~K.}\
  \bibnamefont {Mong}}, \bibinfo {author} {\bibfnamefont {D.~J.}\ \bibnamefont
  {Clarke}}, \bibinfo {author} {\bibfnamefont {J.}~\bibnamefont {Alicea}},
  \bibinfo {author} {\bibfnamefont {N.~H.}\ \bibnamefont {Lindner}}, \ and\
  \bibinfo {author} {\bibfnamefont {P.}~\bibnamefont {Fendley}},\ }\href
  {\doibase 10.1088/1751-8113/47/45/452001} {\bibfield  {journal} {\bibinfo
  {journal} {J. Phys.}\ }\textbf {\bibinfo {volume} {A47}},\ \bibinfo {pages}
  {452001} (\bibinfo {year} {2014})},\ \Eprint {http://arxiv.org/abs/1406.0846}
  {arXiv:1406.0846} \BibitemShut {NoStop}%
\bibitem [{\citenamefont {Rajabpour}\ and\ \citenamefont
  {Cardy}(2008)}]{Rajabpour07}%
  \BibitemOpen
  \bibfield  {author} {\bibinfo {author} {\bibfnamefont {M.~A.}\ \bibnamefont
  {Rajabpour}}\ and\ \bibinfo {author} {\bibfnamefont {J.}~\bibnamefont
  {Cardy}},\ }\href {\doibase 10.1088/1751-8113/40/49/006} {\bibfield
  {journal} {\bibinfo  {journal} {J. Phys.}\ }\textbf {\bibinfo {volume}
  {A40}},\ \bibinfo {pages} {14703} (\bibinfo {year} {2008})},\ \Eprint
  {http://arxiv.org/abs/0708.3772} {arXiv:0708.3772} \BibitemShut {NoStop}%
\bibitem [{\citenamefont {Ikhlef}\ and\ \citenamefont
  {Rajabpour}(2010)}]{Ikhlef2010}%
  \BibitemOpen
  \bibfield  {author} {\bibinfo {author} {\bibfnamefont {Y.}~\bibnamefont
  {Ikhlef}}\ and\ \bibinfo {author} {\bibfnamefont {M.~A.}\ \bibnamefont
  {Rajabpour}},\ }\href {\doibase 10.1088/1751-8113/44/4/042001} {\bibfield
  {journal} {\bibinfo  {journal} {J. Phys. A}\ }\textbf {\bibinfo {volume}
  {44}},\ \bibinfo {pages} {042001} (\bibinfo {year} {2010})}\BibitemShut
  {NoStop}%
\bibitem [{\citenamefont {Ikhlef}\ \emph {et~al.}(2011)\citenamefont {Ikhlef},
  \citenamefont {Fendley},\ and\ \citenamefont {Cardy}}]{Ikhlef11}%
  \BibitemOpen
  \bibfield  {author} {\bibinfo {author} {\bibfnamefont {Y.}~\bibnamefont
  {Ikhlef}}, \bibinfo {author} {\bibfnamefont {P.}~\bibnamefont {Fendley}}, \
  and\ \bibinfo {author} {\bibfnamefont {J.}~\bibnamefont {Cardy}},\ }\href
  {\doibase 10.1103/PhysRevB.84.144201} {\bibfield  {journal} {\bibinfo
  {journal} {Phys. Rev.}\ }\textbf {\bibinfo {volume} {B84}},\ \bibinfo {pages}
  {144201} (\bibinfo {year} {2011})},\ \Eprint {http://arxiv.org/abs/1103.3368}
  {arXiv:1103.3368} \BibitemShut {NoStop}%
\bibitem [{\citenamefont {de~Gier}\ \emph {et~al.}(2013)\citenamefont
  {de~Gier}, \citenamefont {Lee},\ and\ \citenamefont
  {Rasmussen}}]{deGier2012}%
  \BibitemOpen
  \bibfield  {author} {\bibinfo {author} {\bibfnamefont {J.}~\bibnamefont
  {de~Gier}}, \bibinfo {author} {\bibfnamefont {A.}~\bibnamefont {Lee}}, \ and\
  \bibinfo {author} {\bibfnamefont {J.}~\bibnamefont {Rasmussen}},\ }\href
  {\doibase 10.1088/1742-5468/2013/02/P02029} {\bibfield  {journal} {\bibinfo
  {journal} {J. Stat. Mech.}\ }\textbf {\bibinfo {volume} {1302}},\ \bibinfo
  {pages} {P02029} (\bibinfo {year} {2013})},\ \Eprint
  {http://arxiv.org/abs/1210.5036} {arXiv:1210.5036} \BibitemShut {NoStop}%
\bibitem [{\citenamefont {Alam}\ and\ \citenamefont
  {Batchelor}(2014)}]{Alam2014}%
  \BibitemOpen
  \bibfield  {author} {\bibinfo {author} {\bibfnamefont {I.~T.}\ \bibnamefont
  {Alam}}\ and\ \bibinfo {author} {\bibfnamefont {M.~T.}\ \bibnamefont
  {Batchelor}},\ }\href {\doibase 10.1088/1751-8113/47/21/215201} {\bibfield
  {journal} {\bibinfo  {journal} {J. Phys. A}\ }\textbf {\bibinfo {volume}
  {47}},\ \bibinfo {pages} {215201} (\bibinfo {year} {2014})}\BibitemShut
  {NoStop}%
\bibitem [{\citenamefont {Bondesan}\ \emph {et~al.}(2015)\citenamefont
  {Bondesan}, \citenamefont {Dubail}, \citenamefont {Faribault},\ and\
  \citenamefont {Ikhlef}}]{Bondesan14}%
  \BibitemOpen
  \bibfield  {author} {\bibinfo {author} {\bibfnamefont {R.}~\bibnamefont
  {Bondesan}}, \bibinfo {author} {\bibfnamefont {J.}~\bibnamefont {Dubail}},
  \bibinfo {author} {\bibfnamefont {A.}~\bibnamefont {Faribault}}, \ and\
  \bibinfo {author} {\bibfnamefont {Y.}~\bibnamefont {Ikhlef}},\ }\href
  {\doibase 10.1088/1751-8113/48/6/065205} {\bibfield  {journal} {\bibinfo
  {journal} {J. Phys.}\ }\textbf {\bibinfo {volume} {A48}},\ \bibinfo {pages}
  {065205} (\bibinfo {year} {2015})},\ \Eprint {http://arxiv.org/abs/1409.8590}
  {arXiv:1409.8590} \BibitemShut {NoStop}%
\bibitem [{\citenamefont {Ikhlef}\ and\ \citenamefont
  {Weston}(2015)}]{Ikhlef15}%
  \BibitemOpen
  \bibfield  {author} {\bibinfo {author} {\bibfnamefont {Y.}~\bibnamefont
  {Ikhlef}}\ and\ \bibinfo {author} {\bibfnamefont {R.}~\bibnamefont
  {Weston}},\ }\href {\doibase 10.1088/1751-8113/48/29/294001} {\bibfield
  {journal} {\bibinfo  {journal} {J. Phys.}\ }\textbf {\bibinfo {volume}
  {A48}},\ \bibinfo {pages} {294001} (\bibinfo {year} {2015})},\ \Eprint
  {http://arxiv.org/abs/1502.04944} {arXiv:1502.04944} \BibitemShut {NoStop}%
\bibitem [{\citenamefont {Chelkak}\ \emph {et~al.}(2016)\citenamefont
  {Chelkak}, \citenamefont {Glazman},\ and\ \citenamefont
  {Smirnov}}]{Chelkak2016}%
  \BibitemOpen
  \bibfield  {author} {\bibinfo {author} {\bibfnamefont {D.}~\bibnamefont
  {Chelkak}}, \bibinfo {author} {\bibfnamefont {A.}~\bibnamefont {Glazman}}, \
  and\ \bibinfo {author} {\bibfnamefont {S.}~\bibnamefont {Smirnov}},\
  }\href@noop {} {\enquote {\bibinfo {title} {Discrete stress-energy tensor in
  the loop o(n) model},}\ } (\bibinfo {year} {2016}),\ \Eprint
  {http://arxiv.org/abs/1604.06339} {arXiv:1604.06339} \BibitemShut {NoStop}%
\bibitem [{\citenamefont {Ikhlef}\ and\ \citenamefont
  {Weston}(2017)}]{Ikhlef16}%
  \BibitemOpen
  \bibfield  {author} {\bibinfo {author} {\bibfnamefont {Y.}~\bibnamefont
  {Ikhlef}}\ and\ \bibinfo {author} {\bibfnamefont {R.}~\bibnamefont
  {Weston}},\ }\href {\doibase 10.1088/1751-8121/aa63ca} {\bibfield  {journal}
  {\bibinfo  {journal} {J. Phys.}\ }\textbf {\bibinfo {volume} {A50}},\
  \bibinfo {pages} {164003} (\bibinfo {year} {2017})},\ \Eprint
  {http://arxiv.org/abs/1612.03666} {arXiv:1612.03666} \BibitemShut {NoStop}%
\bibitem [{\citenamefont {Bonderson}\ \emph {et~al.}(2019)\citenamefont
  {Bonderson}, \citenamefont {Delaney}, \citenamefont {Galindo}, \citenamefont
  {Rowell}, \citenamefont {Tran},\ and\ \citenamefont {Wang}}]{Bonderson2018}%
  \BibitemOpen
  \bibfield  {author} {\bibinfo {author} {\bibfnamefont {P.}~\bibnamefont
  {Bonderson}}, \bibinfo {author} {\bibfnamefont {C.}~\bibnamefont {Delaney}},
  \bibinfo {author} {\bibfnamefont {C.}~\bibnamefont {Galindo}}, \bibinfo
  {author} {\bibfnamefont {E.~C.}\ \bibnamefont {Rowell}}, \bibinfo {author}
  {\bibfnamefont {A.}~\bibnamefont {Tran}}, \ and\ \bibinfo {author}
  {\bibfnamefont {Z.}~\bibnamefont {Wang}},\ }\href {\doibase
  10.1016/j.jpaa.2018.12.017} {\bibfield  {journal} {\bibinfo  {journal} {J.
  Pure Appl. Algebra}\ }\textbf {\bibinfo {volume} {223}},\ \bibinfo {pages}
  {4065} (\bibinfo {year} {2019})},\ \Eprint {http://arxiv.org/abs/1805.05736}
  {arXiv:1805.05736} \BibitemShut {NoStop}%
\bibitem [{\citenamefont {Feger}\ \emph {et~al.}(2020)\citenamefont {Feger},
  \citenamefont {Kephart},\ and\ \citenamefont {Saskowski}}]{Feger2020}%
  \BibitemOpen
  \bibfield  {author} {\bibinfo {author} {\bibfnamefont {R.}~\bibnamefont
  {Feger}}, \bibinfo {author} {\bibfnamefont {T.~W.}\ \bibnamefont {Kephart}},
  \ and\ \bibinfo {author} {\bibfnamefont {R.~J.}\ \bibnamefont {Saskowski}},\
  }\href {\doibase 10.1016/j.cpc.2020.107490} {\bibfield  {journal} {\bibinfo
  {journal} {Comp. Phys. Communications}\ ,\ \bibinfo {pages} {107490}}
  (\bibinfo {year} {2020})},\ \Eprint {http://arxiv.org/abs/1912.10969}
  {arXiv:1912.10969} \BibitemShut {NoStop}%
\bibitem [{\citenamefont {Kuniba}(1991)}]{Kuniba1991}%
  \BibitemOpen
  \bibfield  {author} {\bibinfo {author} {\bibfnamefont {A.}~\bibnamefont
  {Kuniba}},\ }\href {\doibase 10.1016/0550-3213(91)90495-J} {\bibfield
  {journal} {\bibinfo  {journal} {Nucl. Phys. B}\ }\textbf {\bibinfo {volume}
  {355}},\ \bibinfo {pages} {801} (\bibinfo {year} {1991})}\BibitemShut
  {NoStop}%
\bibitem [{\citenamefont {Izergin}\ and\ \citenamefont
  {Korepin}(1981)}]{Izergin1980}%
  \BibitemOpen
  \bibfield  {author} {\bibinfo {author} {\bibfnamefont {A.~G.}\ \bibnamefont
  {Izergin}}\ and\ \bibinfo {author} {\bibfnamefont {V.~E.}\ \bibnamefont
  {Korepin}},\ }\href {\doibase 10.1007/BF01208496} {\bibfield  {journal}
  {\bibinfo  {journal} {Commun. Math. Phys.}\ }\textbf {\bibinfo {volume}
  {79}},\ \bibinfo {pages} {303} (\bibinfo {year} {1981})}\BibitemShut
  {NoStop}%
\bibitem [{\citenamefont {Fendley}(2001)}]{Fendley2001}%
  \BibitemOpen
  \bibfield  {author} {\bibinfo {author} {\bibfnamefont {P.}~\bibnamefont
  {Fendley}},\ }\href {\doibase 10.1088/1126-6708/2001/05/050} {\bibfield
  {journal} {\bibinfo  {journal} {JHEP}\ }\textbf {\bibinfo {volume} {05}},\
  \bibinfo {pages} {050} (\bibinfo {year} {2001})},\ \Eprint
  {http://arxiv.org/abs/hep-th/0101034} {arXiv:hep-th/0101034} \BibitemShut
  {NoStop}%
\bibitem [{\citenamefont {Date}\ \emph {et~al.}(1987)\citenamefont {Date},
  \citenamefont {Jimbo}, \citenamefont {Kuniba}, \citenamefont {Miwa},\ and\
  \citenamefont {Okado}}]{Date1987}%
  \BibitemOpen
  \bibfield  {author} {\bibinfo {author} {\bibfnamefont {E.}~\bibnamefont
  {Date}}, \bibinfo {author} {\bibfnamefont {M.}~\bibnamefont {Jimbo}},
  \bibinfo {author} {\bibfnamefont {A.}~\bibnamefont {Kuniba}}, \bibinfo
  {author} {\bibfnamefont {T.}~\bibnamefont {Miwa}}, \ and\ \bibinfo {author}
  {\bibfnamefont {M.}~\bibnamefont {Okado}},\ }\href {\doibase
  10.1016/0550-3213(87)90187-8} {\bibfield  {journal} {\bibinfo  {journal}
  {Nucl. Phys.}\ }\textbf {\bibinfo {volume} {B290}},\ \bibinfo {pages} {231}
  (\bibinfo {year} {1987})}\BibitemShut {NoStop}%
\bibitem [{\citenamefont {Kulish}\ \emph {et~al.}(1981)\citenamefont {Kulish},
  \citenamefont {Reshetikhin},\ and\ \citenamefont {Sklyanin}}]{Kulish1981}%
  \BibitemOpen
  \bibfield  {author} {\bibinfo {author} {\bibfnamefont {P.~P.}\ \bibnamefont
  {Kulish}}, \bibinfo {author} {\bibfnamefont {N.~{\relax Yu}.}\ \bibnamefont
  {Reshetikhin}}, \ and\ \bibinfo {author} {\bibfnamefont {E.~K.}\ \bibnamefont
  {Sklyanin}},\ }\href {\doibase 10.1007/BF02285311} {\bibfield  {journal}
  {\bibinfo  {journal} {Lett. Math. Phys.}\ }\textbf {\bibinfo {volume} {5}},\
  \bibinfo {pages} {393} (\bibinfo {year} {1981})}\BibitemShut {NoStop}%
\bibitem [{\citenamefont {Macfarlane}(2001)}]{Macfarlane2001}%
  \BibitemOpen
  \bibfield  {author} {\bibinfo {author} {\bibfnamefont {A.}~\bibnamefont
  {Macfarlane}},\ }\href {\doibase 10.1142/S0217751X01004335} {\bibfield
  {journal} {\bibinfo  {journal} {Int. J. Mod. Phys. A}\ }\textbf {\bibinfo
  {volume} {16}},\ \bibinfo {pages} {3067} (\bibinfo {year} {2001})},\ \Eprint
  {http://arxiv.org/abs/math-ph/0103021} {math-ph/0103021} \BibitemShut
  {NoStop}%
\bibitem [{\citenamefont {Kim}\ \emph {et~al.}(1991)\citenamefont {Kim},
  \citenamefont {Koh},\ and\ \citenamefont {Ma}}]{Kim1990}%
  \BibitemOpen
  \bibfield  {author} {\bibinfo {author} {\bibfnamefont {J.}~\bibnamefont
  {Kim}}, \bibinfo {author} {\bibfnamefont {I.}~\bibnamefont {Koh}}, \ and\
  \bibinfo {author} {\bibfnamefont {Z.-Q.}\ \bibnamefont {Ma}},\ }\href
  {\doibase 10.1063/1.529342} {\bibfield  {journal} {\bibinfo  {journal} {J.
  Math. Phys.}\ }\textbf {\bibinfo {volume} {32}},\ \bibinfo {pages} {845}
  (\bibinfo {year} {1991})}\BibitemShut {NoStop}%
\bibitem [{\citenamefont {Fateev}\ and\ \citenamefont
  {Zamolodchikov}(1982)}]{Fateev82}%
  \BibitemOpen
  \bibfield  {author} {\bibinfo {author} {\bibfnamefont {V.~A.}\ \bibnamefont
  {Fateev}}\ and\ \bibinfo {author} {\bibfnamefont {A.~B.}\ \bibnamefont
  {Zamolodchikov}},\ }\href {\doibase 10.1016/0375-9601(82)90736-8} {\bibfield
  {journal} {\bibinfo  {journal} {Phys. Lett.}\ }\textbf {\bibinfo {volume}
  {A92}},\ \bibinfo {pages} {37} (\bibinfo {year} {1982})}\BibitemShut
  {NoStop}%
\bibitem [{\citenamefont {Tambara}\ and\ \citenamefont
  {Yamagami}(1998)}]{Tambara1998}%
  \BibitemOpen
  \bibfield  {author} {\bibinfo {author} {\bibfnamefont {D.}~\bibnamefont
  {Tambara}}\ and\ \bibinfo {author} {\bibfnamefont {S.}~\bibnamefont
  {Yamagami}},\ }\href {\doibase http://dx.doi.org/10.1006/jabr.1998.7558}
  {\bibfield  {journal} {\bibinfo  {journal} {Journal of Algebra}\ }\textbf
  {\bibinfo {volume} {209}},\ \bibinfo {pages} {692 } (\bibinfo {year}
  {1998})}\BibitemShut {NoStop}%
\bibitem [{\citenamefont {Zamolodchikov}\ and\ \citenamefont
  {Fateev}(1985)}]{Zamolodchikov85}%
  \BibitemOpen
  \bibfield  {author} {\bibinfo {author} {\bibfnamefont {A.~B.}\ \bibnamefont
  {Zamolodchikov}}\ and\ \bibinfo {author} {\bibfnamefont {V.}~\bibnamefont
  {Fateev}},\ }\href {http://www.jetp.ac.ru/cgi-bin/e/index/r/89/2/p380?a=list}
  {\bibfield  {journal} {\bibinfo  {journal} {Zh. Eksp. Teor. Fiz.}\ }\textbf
  {\bibinfo {volume} {89}},\ \bibinfo {pages} {380} (\bibinfo {year} {1985})},\
  \bibinfo {note} {[Eng. transl. JETP, {\bf 62}, 215--225]}\BibitemShut
  {NoStop}%
\bibitem [{\citenamefont {Babichenko}(2003)}]{Babichenko2002}%
  \BibitemOpen
  \bibfield  {author} {\bibinfo {author} {\bibfnamefont {A.}~\bibnamefont
  {Babichenko}},\ }\href {\doibase 10.1016/S0370-2693(02)03268-9} {\bibfield
  {journal} {\bibinfo  {journal} {Phys. Lett. B}\ }\textbf {\bibinfo {volume}
  {554}},\ \bibinfo {pages} {96} (\bibinfo {year} {2003})},\ \Eprint
  {http://arxiv.org/abs/hep-th/0211114} {arXiv:hep-th/0211114} \BibitemShut
  {NoStop}%
\bibitem [{\citenamefont {Baxter}(1978)}]{Baxter1978}%
  \BibitemOpen
  \bibfield  {author} {\bibinfo {author} {\bibfnamefont {R.~J.}\ \bibnamefont
  {Baxter}},\ }\href {\doibase 10.1098/rsta.1978.0062} {\bibfield  {journal}
  {\bibinfo  {journal} {Phil. Trans. Roy. Soc. Lond. A}\ }\textbf {\bibinfo
  {volume} {289}},\ \bibinfo {pages} {315} (\bibinfo {year}
  {1978})}\BibitemShut {NoStop}%
\bibitem [{\citenamefont {Zhou}\ and\ \citenamefont
  {Batchelor}(1997)}]{Zhou1997}%
  \BibitemOpen
  \bibfield  {author} {\bibinfo {author} {\bibfnamefont {Y.}~\bibnamefont
  {Zhou}}\ and\ \bibinfo {author} {\bibfnamefont {M.}~\bibnamefont
  {Batchelor}},\ }\href {\doibase 10.1016/s0550-3213(96)00654-2} {\bibfield
  {journal} {\bibinfo  {journal} {Nuclear Physics B}\ }\textbf {\bibinfo
  {volume} {485}},\ \bibinfo {pages} {646} (\bibinfo {year}
  {1997})}\BibitemShut {NoStop}%
\bibitem [{\citenamefont {Muger}(2003)}]{Muger2003}%
  \BibitemOpen
  \bibfield  {author} {\bibinfo {author} {\bibfnamefont {M.}~\bibnamefont
  {Muger}},\ }\href {\doibase http://dx.doi.org/10.1016/S0022-4049(02)00248-7}
  {\bibfield  {journal} {\bibinfo  {journal} {Journal of Pure and Applied
  Algebra}\ }\textbf {\bibinfo {volume} {180}},\ \bibinfo {pages} {159 }
  (\bibinfo {year} {2003})}\BibitemShut {NoStop}%
\bibitem [{\citenamefont {{Aasen}}\ and\ \citenamefont
  {Mong}(2016)}]{Aasen2020b}%
  \BibitemOpen
  \bibfield  {author} {\bibinfo {author} {\bibfnamefont {D.}~\bibnamefont
  {{Aasen}}}\ and\ \bibinfo {author} {\bibfnamefont {R.}~\bibnamefont {Mong}},\
  }\href@noop {} {\  (\bibinfo {year} {2016})},\ \bibinfo {note}
  {unpublished}\BibitemShut {NoStop}%
\bibitem [{\citenamefont {Hong}\ \emph {et~al.}(2008)\citenamefont {Hong},
  \citenamefont {Rowell},\ and\ \citenamefont {Wang}}]{Hong2008}%
  \BibitemOpen
  \bibfield  {author} {\bibinfo {author} {\bibfnamefont {S.-M.}\ \bibnamefont
  {Hong}}, \bibinfo {author} {\bibfnamefont {E.}~\bibnamefont {Rowell}}, \ and\
  \bibinfo {author} {\bibfnamefont {Z.}~\bibnamefont {Wang}},\ }\href {\doibase
  10.1142/s0219199708003162} {\bibfield  {journal} {\bibinfo  {journal}
  {Communications in Contemporary Mathematics}\ }\textbf {\bibinfo {volume}
  {10}},\ \bibinfo {pages} {1049} (\bibinfo {year} {2008})},\ \Eprint
  {http://arxiv.org/abs/0710.5761} {arXiv:0710.5761} \BibitemShut {NoStop}%
\bibitem [{\citenamefont {Gould}\ and\ \citenamefont
  {Zhang}(2002)}]{Gould2002}%
  \BibitemOpen
  \bibfield  {author} {\bibinfo {author} {\bibfnamefont {M.~D.}\ \bibnamefont
  {Gould}}\ and\ \bibinfo {author} {\bibfnamefont {Y.-Z.}\ \bibnamefont
  {Zhang}},\ }\href {\doibase 10.1142/S0217979202011901} {\bibfield  {journal}
  {\bibinfo  {journal} {Int. J. Mod. Phys. B}\ }\textbf {\bibinfo {volume}
  {16}},\ \bibinfo {pages} {2145} (\bibinfo {year} {2002})},\ \Eprint
  {http://arxiv.org/abs/hep-th/0205071} {arXiv:hep-th/0205071} \BibitemShut
  {NoStop}%
\bibitem [{\citenamefont {Costello}\ \emph {et~al.}(2018)\citenamefont
  {Costello}, \citenamefont {Witten},\ and\ \citenamefont
  {Yamazaki}}]{Costello2018}%
  \BibitemOpen
  \bibfield  {author} {\bibinfo {author} {\bibfnamefont {K.}~\bibnamefont
  {Costello}}, \bibinfo {author} {\bibfnamefont {E.}~\bibnamefont {Witten}}, \
  and\ \bibinfo {author} {\bibfnamefont {M.}~\bibnamefont {Yamazaki}},\ }\href
  {\doibase 10.4310/iccm.2018.v6.n1.a6} {\bibfield  {journal} {\bibinfo
  {journal} {Notices of the International Congress of Chinese Mathematicians}\
  }\textbf {\bibinfo {volume} {6}},\ \bibinfo {pages} {46} (\bibinfo {year}
  {2018})},\ \Eprint {http://arxiv.org/abs/1709.09993} {arXiv:1709.09993}
  \BibitemShut {NoStop}%
\end{thebibliography}%
\bibliographystyle{apsrev4-1}

\end{document}